\documentclass{emulateapj}

\usepackage{aas_macros}
\newcommand{\Msol}{{\>{\rm M}_\odot}} 
\newcommand{\hMsol}{{\>h^{-1}\rm M}_\odot}             
\newcommand{\hMpc}{{\>h^{-1}\rm  Mpc}}
\newcommand{\hkpc}{{\>h^{-1}\rm  kpc}}  
\newcommand{\keV}{\>{\rm keV}}
\newcommand{\kms}{\>{\rm   km\,s^{-1}}}  

\usepackage{graphicx} \usepackage{amssymb} \usepackage{amsmath}

\shorttitle{The velocity  function in the local  environment}
\shortauthors{Zavala et al.}

\begin{document}

\title{The velocity
  function in the local environment from $\Lambda$CDM and $\Lambda$WDM
  constrained  simulations}  

\author{J.     Zavala\altaffilmark{1,2},    Y.     P.    Jing\altaffilmark{1},
  A.       Faltenbacher\altaffilmark{2,1},      G.      Yepes\altaffilmark{3},
  Y.     Hoffman\altaffilmark{4},    S.     Gottl\"ober\altaffilmark{5}    and
  B. Catinella\altaffilmark{2}}
\altaffiltext{1}{MPA/SHAO Joint Center for Astrophysical Cosmology at Shanghai
  Astronomical   Observatory,  Nandan   Road  80,   Shanghai   200030,  China}
\altaffiltext{2}{Max-Planck-Institute             for            Astrophysics,
  Karl-Schwarzschild-Str.       1,       D-85741      Garching,       Germany}
\altaffiltext{3}{Grupo de Astrof\'{i}sica, Universidad Aut\'{o}noma de Madrid,
  E-28049,   Spain}  \altaffiltext{4}{Racah   Institute  of   Physics,  Hebrew
  University,  Jerusalem  91904, Israel}  \altaffiltext{5}{Astrophysikalisches
  Institut Potsdam, An der Sternwarte 16, D-14482 Potsdam, Germany}

\begin{abstract}
Using  constrained simulations  of the  local Universe  for generic  cold dark
matter and for 1$\keV$ warm dark  matter, we investigate the difference in the
abundance of dark matter halos in the local environment. We find that the mass
function within $20\hMpc$ of the Local  Group is $\sim2$ times larger than the
universal mass  function in the $10^9-10^{13}\hMsol$ mass  range. Imposing the
field of view  of the on-going HI blind survey ALFALFA  in our simulations, we
predict that the  velocity function in the Virgo-direction  region exceeds the
universal velocity function by a factor of 3.  Furthermore, employing a scheme
to translate  the halo velocity function  into a galaxy  velocity function, we
compare  the simulation  results  with a  sample  of galaxies  from the  early
catalog  release  of  ALFALFA.  We  find  that  our  simulations are  able  to
reproduce the velocity  function in the $80-300\kms$ velocity  range, having a
value  $\sim10$ times  larger  than  the universal  velocity  function in  the
Virgo-direction  region.  In the  low velocity  regime, $35-80\kms$,  the warm
dark  matter simulation  reproduces the  observed flattening  of  the velocity
function. On  the contrary,  the simulation with  cold dark matter  predicts a
steep   rise  in  the   velocity  function   towards  lower   velocities;  for
$V_{max}=35\kms$,  it forecasts  $\sim10$  times more  sources  than the  ones
observed. If confirmed by the  complete ALFALFA survey, our results indicate a
potential problem  for the cold dark  matter paradigm or  for the conventional
assumptions about energetic feedback in dwarf galaxies.
\end{abstract}

\keywords{Cosmology: dark matter, galaxies: Local Group}

\section{Introduction}
Despite the success of the $\Lambda$CDM paradigm in describing the large scale
structure of  the Universe, as has  been attested by  improved measurements of
the     CMB    and     the    large     scale    clustering     of    galaxies
\citep{Seljak-Slosar-McDonald-06,Komatsu-08},  some potential  problems remain
for the  model at smaller scales.  One  of these challenges is  related to the
prediction  of the  abundance  of low  mass  galaxies.  Numerical  simulations
within the  $\Lambda$CDM model predict  many more dark matter  subhalos inside
galactic-sized  host  halos  than  the  actual number  of  observed  satellite
galaxies             around              the             Milky             Way
\citep{Klypin-99,Moore-99,Diemand-Kuhlen-Madau-07,Springel-08}.            This
discrepancy  has been known  as the  ``missing satellite  problem''.  Although
some astrophysical processes can be  responsible for the suppression of galaxy
formation  in  small mass  halos  and  lead to  the  solution  of the  problem
\citep{Somerville-02,Benson-02a,Benson-02b,Gnedin-Kravtsov-06,Koposov-09},   it
is also possible  to alleviate the discrepancy by  adopting alternative models
with a different type of dark matter.

Warm  dark  matter (WDM)  models,  for  instance,  predict significantly  less
substructure                            within                           halos
\citep{Coln-Avila-Reese-Valenzuela-00,Bode-Ostriker-Turok-01}.     The    free
streaming  length  for  dark  matter  particles  depends  on  their  intrinsic
properties  and  establishes a  cut-off  scale  for  halo masses  below  which
primordial density perturbations  are wiped out.  In the  case of neutralinos,
one of the favorite candidates  for CDM models, with $m_{\chi}\sim100$GeV, the
scale  is roughly  $10^{-6}\Msol$  \citep{Diemand-Moore-Stadel-05}.  For
WDM   candidates,  such  as   gravitinos  with   $m_{\tilde{G}}\sim1$KeV  (see
\citealt{Steffen-06} for  a recent review on gravitino  cosmology), this scale
is roughly  $10^{10}\Msol$ (see section  2).  Therefore, using  the same
cosmological  parameters but  different  cut-off scales  results in  different
predictions  for  the  halo mass  function  (MF).  At  the  low mass  end  the
$\Lambda$WDM  mass function  is expected  to be  significantly lower  than the
$\Lambda$CDM mass  function.  Above the  cut-off scale the  differences vanish
\citep[e.g.][]{Bode-Ostriker-Turok-01,Barkana-Haiman-Ostriker-01}.

Since  for  high  density   environments  both  models  provide,  conceptually
different, solutions for the missing satellite problem it is also important to
test their predictions for isolated  systems in low density environments which
are  not affected  by processes  such as  ram-pressure or  tidal  striping.  A
comparison  between $\Lambda$WDM  and  $\Lambda$CDM is  then  free from  these
astrophysical phenomena and conclusions drawn from it are more closely related
to the  nature of dark matter itself.  Such an approach has  been employed for
example by \citet{Blanton-Geha-West-08}.

Even though  low mass galaxies  are difficult to  detect, surveys such  as the
ongoing Arecibo  Legacy Fast ALFA (ALFALFA)  survey \citep{Giovanelli-05a} are
promising to make such comparison.  ALFALFA is exploring the local HI Universe
with a  sky coverage of  $7000$ deg$^2$ and  it is designed to  detect objects
with  masses  as  low  as  $3\times10^{7}\Msol$ at  a  Virgo-cluster  distance
($11.7\hMpc$) \citep{Giovanelli-07}. The survey is divided into two regions on
the sky.  One within  the solid angle  $07^h30^m<$R.A.(J2000.0)$<16^h30^m$ and
$0\degr<$DEC(J2000.0)$<+36\degr$,  which  includes   the  Virgo  cluster,  and
another     region     within    the     same     declination    range     and
$22^h<$R.A.(J2000.0)$<03^h$.  In the reminder  of this  work, we  will loosely
refer to  these two regions,  additionally constrained to distances  less than
$20\hMpc$, as the  ``Virgo-direction region (VdR)'' and ``anti-Virgo-direction
region (aVdR)'' respectively. We note that the catalogs released by ALFALFA so
far in  these two regions effectively correspond  to a high and  a low density
environment. The HI sources detected  by ALFALFA are typically associated with
star forming  disk galaxies, spirals  in rich clusters  are less likely  to be
detected since these galaxies might be HI deficient \citep{Solanes-01}.

In order  to make predictions on the  abundance of low mass  galaxies based on
the  $\Lambda$CDM and  $\Lambda$WDM models  for our  local environment,  it is
advantageous  to  use constrained  simulations  (CSs)  of  the local  universe
\citep[e.g.][]{1998ApJ...492..439B,Mathis-02,Klypin-03}.    These  simulations
are constructed to  reproduce the gross features of  the nearby Universe, such
as the  Local Supercluster  and the  Virgo cluster.  This  can be  achieved by
setting the initial conditions  of the simulations as constrained realizations
of Gaussian  fields, where  actual observational data  are used to  impose the
constraints. By  using such simulations,  biases which are present  because of
the particular structure of the  local environment, are minimized and a better
comparison between simulations and observations can be achieved.

Although the abundance of dark matter  halos is usually quantified by the mass
function, which describes  the number density of halos as  a function of mass,
usually   presented    either   in    a   differential   or    integral   form
\citep[e.g.][]{Reed-07,Lukic-07},  an  alternative  approach  is  to  use  the
velocity  function, similarly  defined as  the number  density of  halos  as a
function  of  maximum   circular  velocity  \citep[e.g.][]{Gonzalez-00}.   The
advantage  of the  velocity function  (VF)  is that  it can  be compared  more
directly  with observational  results  since it  avoids  the more  complicated
problem of relating  dark matter halo masses to  galaxy luminosities. Instead,
the velocity  function of  galaxies can be  theoretically estimated  using the
velocity function  of halos by incorporating  a model of  baryonic infall; the
only processes  that affect  the velocity function  are those that  modify the
gravitational  potential of  galactic systems.   In the  present work  we show
analysis  of both, the  mass and  velocity functions,  but concentrate  on the
latter for comparisons with the ALFALFA survey.

The  objective of  our  paper is  to  run a  $\Lambda$CDM  and a  $\Lambda$WDM
simulation of  the local  environment where the  same set of  constraints have
been  imposed and to  compare the  abundance and  distribution of  dark matter
halos for the  two cosmologies.  In addition, the results  are used to predict
the abundance  of HI sources  that will be  detected by the on-going  HI blind
survey ALFALFA.

The paper  is organized as follows.  In  section 2 we describe  the setting of
the simulations. The definition of the coordinate system that we use to impose
the field of view of ALFALFA in  the CSs is described in section 3. In section
4 we present results  on the abundance of dark matter halos.   In section 5 we
analyze  in particular the  abundance for  the restricted  Virgo-direction and
anti-Virgo-direction  regions cataloged  by  the ALFALFA  survey  to date  and
present predictions  on the velocity function  of HI sources.  The summary and
conclusions of our work are given in section 6.

\section{Constrained simulations of the local environment}
\label{sec:simulation}
\begin{figure}
\centering                  \includegraphics[height=14cm,width=14cm]{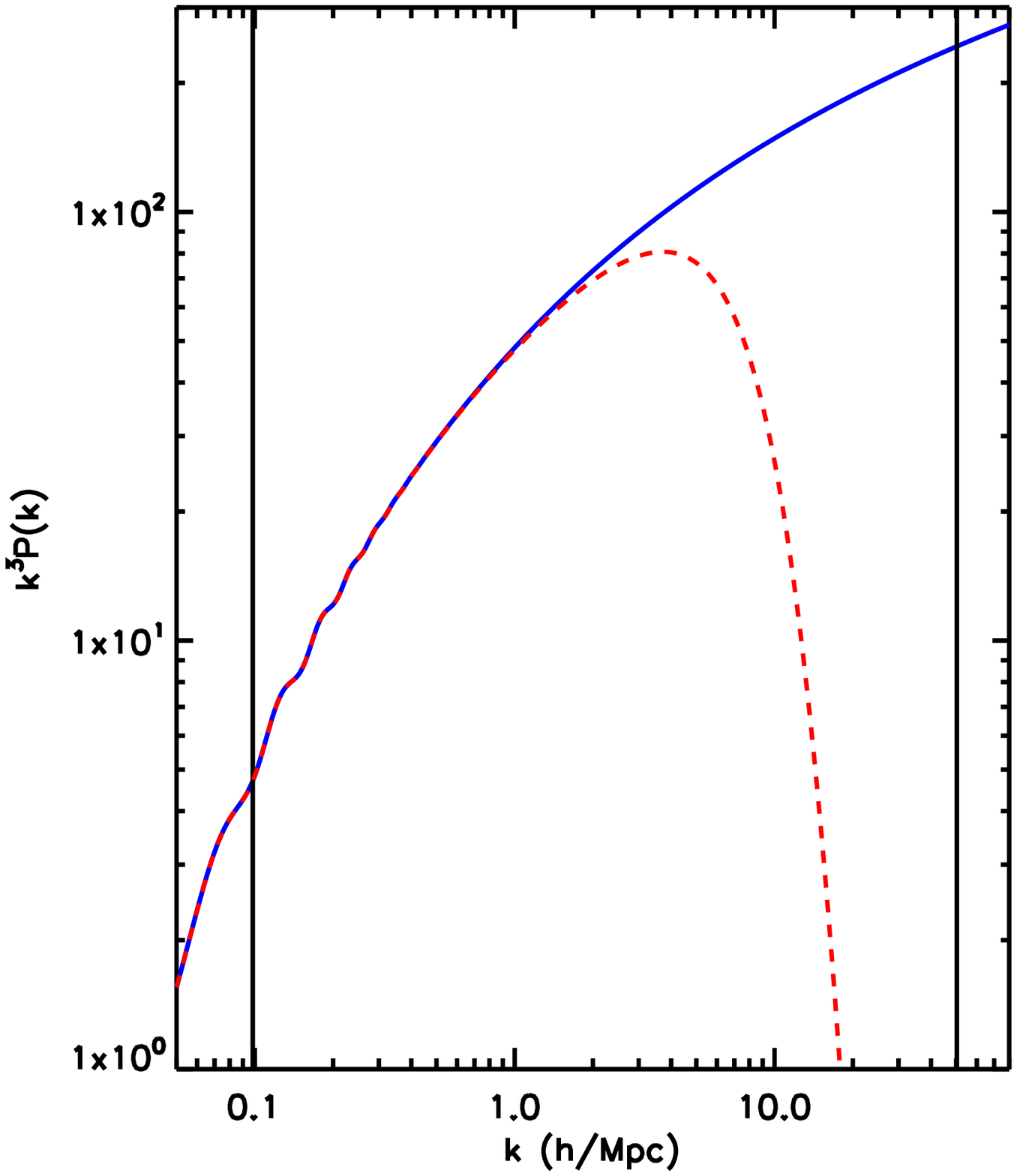}
\caption{Unconstrained linear  power spectrum at z=0 for  the $\Lambda$CDM and
  $\Lambda$WDM ($m_{WDM}=1\keV$) cosmologies (blue and red respectively).  The
  associated Nyquist frequency and fundamental mode for our set of simulations
  are represented  with vertical lines  to the right  and left in  the figure,
  respectively.}
\label{Power}
\end{figure}

We chose the cosmological parameters for our simulations to be consistent with
the     WMAP    3-year     results     \citep{Spergel-07}:    $\Omega_m=0.24$,
$\Omega_{\Lambda}=0.76$,   $H_0=100h\rm\,km\,s^{-1}Mpc^{-1}$   with  $h=0.73$,
$n=0.95$   and  $\sigma_8=0.75$.    For  both   cosmologies   the  theoretical
(unconstrained) linear power  spectrum at $z=0$ is shown  in Fig.  \ref{Power}
(blue for CDM and red for WDM).  The CDM power spectrum was computed from a
Boltzmann code by W. Hu and was kindly provided to us.

\subsection{$\Lambda$WDM simulation settings}
\label{sec:WDM_settings}

The WDM power spectrum was computed  by rescaling the CDM power spectrum using
a fitting function  that approximates the transfer function  for a thermal WDM
particle with $m_{WDM}=1\keV$.  The filtering scale (or free-streaming length)
for  this WDM  particle  mass is  $350\hkpc$,  or about
$\sim1.1\times10^{10}\hMsol$ 
for the filtering  mass (following the definitions given by
\citealt{Bode-Ostriker-Turok-01}).    We   used   the  fitting   function   from
\citet{Viel-05} (see their eqs. 5-7) which is very similar to the one given by
\citet{Bode-Ostriker-Turok-01},  however, according  to the  authors, it  is a
better fitting formula for Boltzmann codes.

As  stated by  recent Lyman-$\alpha$  forest, CMB  and galaxy  clustering data
\citep{Viel-06,Seljak-06}, see also \citet{Miranda-Maccio-07}, the lower limit
for the mass of thermal WDM candidates, for the case where WDM is the dominant
form  of  dark matter,  has  been reported  to  be  $m_{WDM}\sim2\keV$ at  the
$2\sigma$  level.   However,  these   estimates  for  thermal  relics  may  be
contaminated by  systematic errors (see for example  \citet{Boyarsky-08} for a
discussion  on  the  complications  associated to  the  Lyman-$\alpha$  forest
method).  For instance, the constraint given in \cite{Viel-06} would reduce to
$m_{WDM}\geq0.9\keV$ if the highest redshift bins of the Lyman-$\alpha$ forest
data are rejected  from the analysis and only the more  reliable data based on
$z<3.2$     is    taken     into    account.      In    a     recent    paper,
\citet{Boyarsky-Ruchayskiy-Iakubovskyi-08}  revisit the  lower  bounds on  the
mass  of  WDM  particles  and  find  $m_{WDM}\geq1.7\keV$  ($2\sigma$  level).
However, after  discussing with detail  the systematic uncertainties  in their
method they  conclude that the mass  bounds are reliable  within $\sim30\%$ of
uncertainty.  All  these analysis put  our choice of $m_{WDM}=1\keV$  close to
the most recent lower bound, but it is a choice that is still not ruled out.

The  velocity dispersion  of  the WDM  particles  needs, in  principle, to  be
introduced in the  initial conditions for the simulations  since after all, it
is what causes  the smearing of small scale  primordial perturbations. This is
usually done by introducing a random velocity field according to a Fermi-Dirac
distribution              function              \citep{Bode-Ostriker-Turok-01,
  Coln-Valenzuela-Avila-Reese-08}.  However, if this velocity dispersion field
is introduced  at random, a certain  degree of white noise  is generated (shot
noise)  due  to the  finite  number  of  simulation particles  which  produces
spurious   small   scale  power   in   the   WDM   spectrum  (see   Fig.1   of
\citealt{Coln-Valenzuela-Avila-Reese-08}).  The amplitude of the rms velocity of
the random component to be added depends on the nature of the WDM particle and
on the redshift  of the initial conditions.  For thermal  relics, it is larger
for higher redshifts and smaller particle masses.  For $z=50$ the rms velocity
is  $\sim2.2\kms$   for  $m_{WDM}=1\keV$  (following  the   formula  given  by
\citealt{Bode-Ostriker-Turok-01}),  far  lower than  the  velocities induced  by
gravitational  collapse of structures  having scales  larger than  the minimum
scale we can resolve within  our simulations.  Therefore we do not incorporate
any random velocity field into our initial conditions.

To back our  approach we computed the comoving Jeans  length associated with a
velocity dispersion of $2.2\kms$.  We found that the time to build up pressure
support against  gravitational collapse  is similar to  the collapse  time for
comoving scales of $\sim33 \hkpc$,  corresponding to a Jeans mass of
$\sim10^6\hMsol$.  
Differences in the  structure formation are only expected
for masses  of the order  and below  the Jeans mass,  which is lower  than the
particle  mass in our  simulations (see  below). Thus,  the effect  of thermal
velocities is too small  to have a notable effect on the  abundance of low mass
halos in our simulations.  Still, they  could have an impact in the inner part
of halos \citep{Coln-Valenzuela-Avila-Reese-08}. However, since the shot noise
produced by their introduction has a  large spurious effect, we decided not to
include them in  our simulations.  For the purposes of this  work, this has no
consequences in our analysis.
%%
%%Fig. 1
%%

\subsection{Constrained simulations}
\label{sec:CS}

The initial conditions  for the constrained simulations were  set up using the
\citep{Hoffman-Ribak-91}  algorithm of  constrained  realizations of  Gaussian
random  fields.   Two   types  of  data  sets  are  used   as  input  for  the
algorithm. The first  data set is made of radial  velocities of galaxies drawn
from   the   catalogs:  MARK   III   \citep{Willick-97},  surface   brightness
fluctuations   \citep{Tonry-01}   and   the   Catalog   of   Nearby   Galaxies
\citep{Karachentsev-04}.  Present epoch  peculiar velocities are less affected
by non-linear effects  and are therefore imposed as  linear constraints on the
primordial perturbation  field \citep{Zaroubi-Hoffman-Dekel-99}. This approach
follows   the   CSs   performed  by   \citet{Kravtsov-Klypin-Hoffman-02}   and
\citet{Klypin-03}. The other  data set is obtained from  the catalog of nearby
X-ray selected  clusters of galaxies  \citep{Reiprich-Bohringer-02}.  Assuming
the spherical top-hat  model and using the virial parameters  of a cluster the
linear  over-density  of  the   cluster  is  derived.   The  estimated  linear
overdensity  is  imposed  on  the  virial  mass scale  of  the  cluster  as  a
constraint. The density and velocity fields on scales larger than 5$\hMpc$ are
strongly constrained by the imposed data.

Using the above  initial conditions, we carried out  the simulations using the
code GADGET-2  \citep{Springel-05}.  Both simulations follow  the evolution of
$1024^3$  dark  matter  particles from  $z=50$  to  $z=0$  in  a box  of  size
$L=64\hMpc$.   The  associated  Nyquist  frequency and  fundamental  mode  are
represented  with  vertical  lines  in  Fig.  \ref{Power}.   The  dark  matter
particle  mass  is  $m_{DM}=1.63\times10^{7}\hMsol$. The  simulations
were started  with a fixed  comoving softening length  (Plummer-equivalent) of
$\epsilon=1.6\hkpc$.  Once  the corresponding physical  comoving softening
reached $\epsilon=0.8\hkpc$, it was kept constant at this value.

%%
%%Fig.2
%%

Fig.~\ref{Projection} displays  the projected dark matter  distribution of the
CSs at  $z=0$ for the  $\Lambda$CDM and the $\Lambda$WDM  models respectively.
The highest and lowest concentrations  are colored red and black respectively.
Both panels show the matter distribution within a slice that is $8\hMpc$ thick
projected onto the X-Y plane and centered on $Z=24\hMpc$.  The figure captures
the Local Supercluster  (LSC) which is the filamentary  structure crossing the
image plane horizontally.  It is the most prominent feature.  The locations of
the virtual Local Group (LG)\footnote{In what follows, the LG refers to a pair
  of halos associated with the MW  and M31 galaxies} and the Virgo cluster are
marked as  well.  A  description of the  identification of these  objects will
follow  below.  A  visual  inspection of  the  two figures  already reveals  a
deficit of small scale structure in the $\Lambda$WDM simulation, as expected.

Halos      in       the      simulations      were       identified      using
AHF\footnote{http://www.aip.de/People/AKnebe/AMIGA/},   AMIGA's   halo  finder
\citep{Knollman-Knebe-09}.  AHF identifies halos as local density maxima in an
adaptively smoothed density field using  a hierarchy of grids and a refinement
criterion.  The latter was chosen so that a grid cell is subdivided until each
subsection  contains less  than 5  particles.  In  this way,  the size  of the
smallest  cells is  comparable to  the  force resolution  of our  simulations.
Halos,  and/or  subhalos,   are  formed  by  groups  of   particles  that  are
gravitationally bound  to a given  density peak.  For  each halo in  the final
catalog,  AHF provides  a  list  of internal  properties,  the most  important
quantities for the  current study are: the viral  radius $r_{vir}$, defined as
the       radius       that       contains      a       mean       overdensity
$\bar{\rho}(r_{vir})=\Delta\rho_{crit}$,  where $\rho_{crit}$ is  the critical
density and we have chosen $\Delta\sim94$ at $z=0$, a value computed according
to the  spherical collapse model for  our cosmological parameters\footnote{For
  $\Omega_m+\Omega_{\Lambda}=1$,              $\Delta\approx178\Omega_m^{0.45}$
  \citep[e.g.][]{Eke-Navarro-Frenk-98}};   the   corresponding   virial   mass
$M_{vir}$; the  maximum rotational velocity $V_{max}$; and  the spin parameter
$\lambda=J\sqrt{\vert E\vert}~/~GM_{vir}^{5/2}$  \citep{Peebles-69}, where $J$
and  $E$ are  the total  angular momentum  and energy  of the  halo.  Subhalos
identified within each halo were removed from the catalog; for the reminder of
our analysis we will use main halos only.
\section{Definition of the coordinate system}
\label{sec:Coordinates}
%%
%%Fig.3
%%

The  significance of  our study  relies on  a proper  simulation of  the local
environment.  However,  the constraints imposed  on the initial  conditions of
the simulations leave  some freedom for the evolution,  mainly on small scales
($<5\hMpc$).  This can partly be accounted  for by a careful adjustment of the
coordinate  system. We  have  done so  using  the following  steps.  First  we
identified an appropriate LG within the  CSs.  To that purpose we followed the
criteria described in \cite{Maccio-Governato-Horellou-05} (see also table 2 of
\citealt{Martinez-Vaquero-Yepes-Hoffman-07}) to  identify LG candidates within
the  CSs.   These  are  i)  match  in mass,  proximity  and  kinematics  of  a
MW-M31-halo-like  binary system,  i.e., we  look for  a pair  of halos  with a
maximum rotational velocity, $V_{max}$,  between $125\kms$ and $270\kms$, with
a distance  between the  pair $\lesssim1 \hMpc$  and with a  negative relative
velocity; ii)  absence of a massive  near-by halo, i.e., no  halos with masses
larger than  the members  of the pair  within a  radius of $2\hMpc$;  and iii)
presence  of a Virgo-like  halo, $500<V_{max}<1500  \kms$, at  the appropriate
distance, $5-12\hMpc$.

In addition, we  favor LG candidates which are located close  to the center of
the  simulation box  since the  constrained  initial conditions  place the  LG
progenitor exactly at the center.  Subsequent dynamical evolution may displace
the entire environment by some Mpcs.

%%
%%Table 1
%%

\begin{figure*}
\centering                      \includegraphics[width=0.49\hsize]{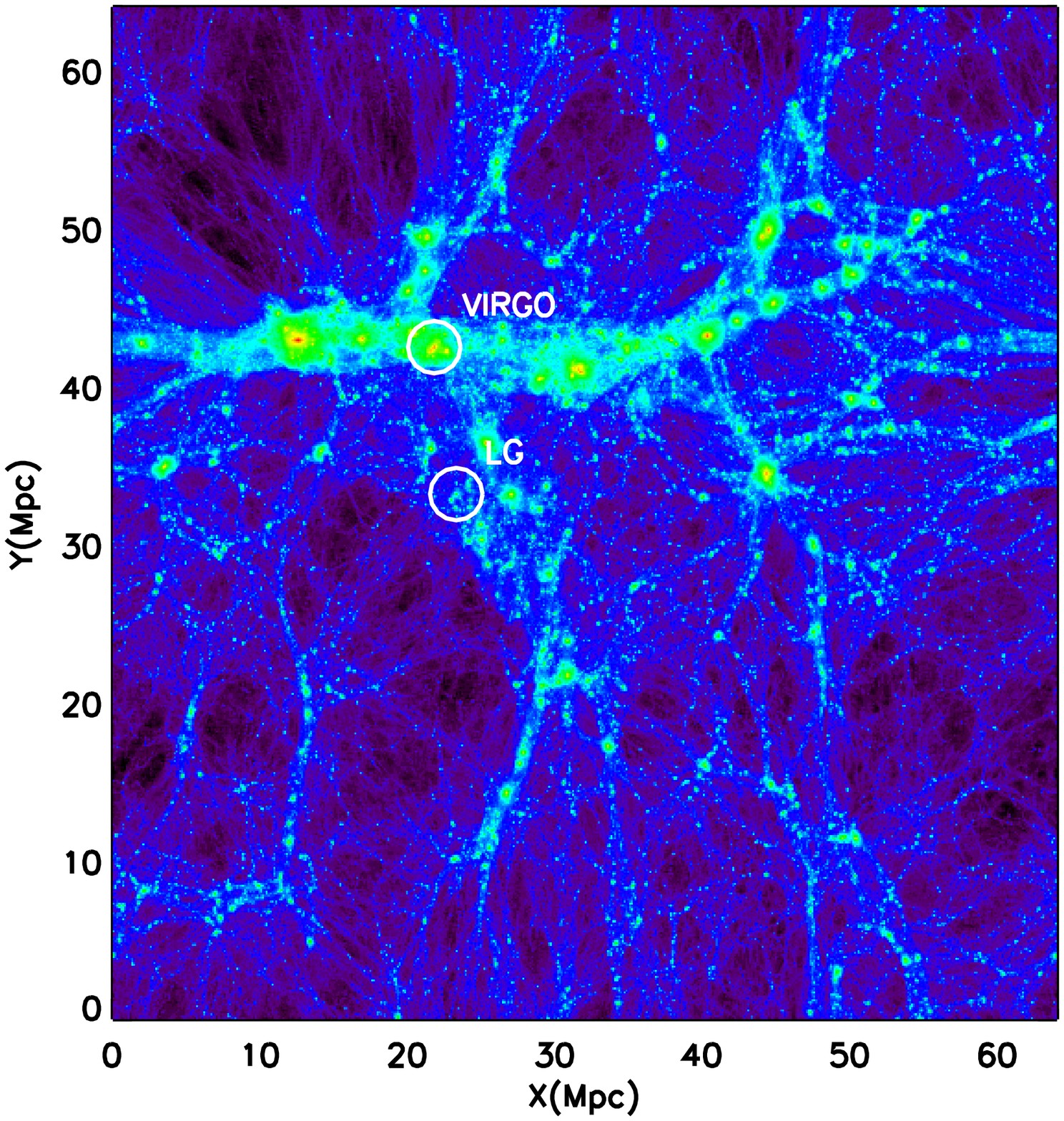}
\includegraphics[width=0.49\hsize]{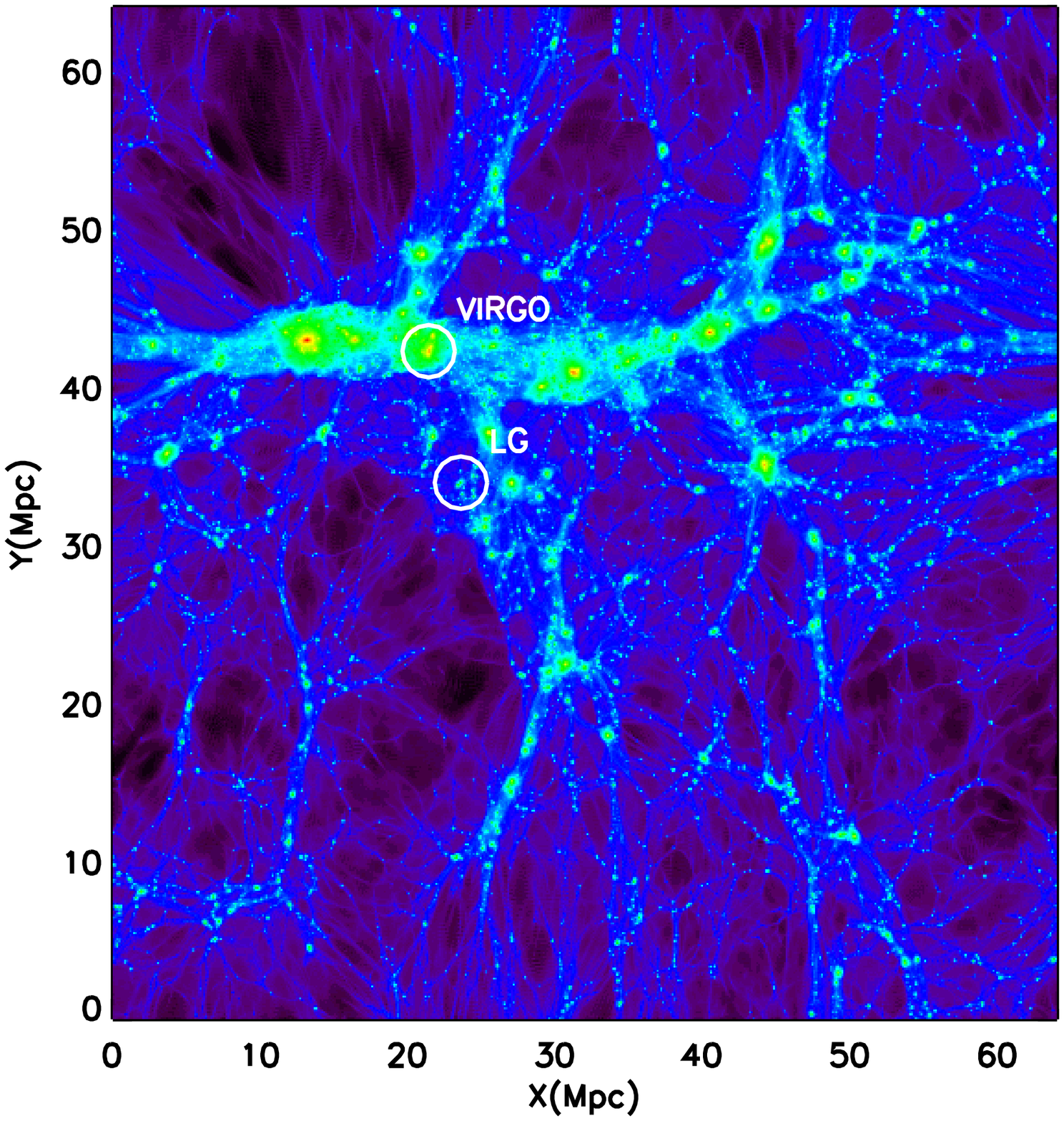}
\caption{\label{Projection}   {\it   Left   panel:}  projected   dark   matter
  distribution  for the  $\Lambda$CDM  constrained simulation  at $z=0$.   The
  slice has a thickness of 8$\hMpc$  and it is centered at $Z=24\hMpc$ (in Box
  coordinates).  It  encompasses part of  the supergalactic plane.   The Local
  Supercluster is the filament  crossing the image horizontally. The locations
  of the  Local Group and  the Virgo cluster  are marked in the  figure.  {\it
    Right panel:}  same  as on the left for  the $\Lambda$WDM
  simulation.}
\end{figure*}

Slices containing the best LG candidates for  the CDM and WDM run are shown in
Fig.~\ref{Projection}  where we have  marked the  location of  the LG  and the
Virgo cluster.  If not stated otherwise, ``LG'' and ``Virgo'' denote these best
possible candidates. Table  1 contains some properties of  the main objects in
our   simulations:  the   LG,  and   the  clusters   Virgo  and   Fornax  (see
below). Observed estimates for these properties are also given in the table.

Based on the locations  of the LG, Virgo and the local  environment we now aim
to  define a  supergalactic coordinate  system  (\citet{deVaucouleurs-91}; see
also \citealt{Lahav-00}).  We will use  such coordinate system as the basis to
identify the  ALFALFA regions in the  (simulated) sky. To define  it, we first
assume that the  equatorial plane of the supergalactic  coordinate system lies
in the  supergalactic plane  (SGP), which is  spanned by  the LG and  the LSC.
Thus, besides Virgo,  we need to find another cluster belonging  to the LSC to
mathematically define the LSC.  As revealed by Fig.~\ref{Projection}, there is
a prominent  cluster on the right hand  side of Virgo.  Its  location and mass
closely    resemble   those    of    the   observed    Ursa   Major    cluster
($1.6\times10^{13}\hMsol$ at a distance of $\sim11.1\hMpc$ from the
LG).   Albeit this choice  seems natural  we have  also tested  other clusters
within the simulated LSC to  define alternative coordinate systems.  The final
results turned out to be almost  independent of this choice. Using this SGP we
construct an  orthonormal vector basis  with the origin  placed at the  MW and
rotate it until  the simulated Virgo cluster is located  at the same longitude
of  the   observed  one,   i.e.,  at  a   supergalactic  longitude   (SGL)  of
$102.45\degr$. Finally, we tilt the  plane to achieve a supergalactic latitude
(SGB) of $2.84\degr$  to match the one that is  observed. With this procedure,
the simulated  Virgo is located at  the same angular position  as the observed
one.   Below we  will introduce  an alternative  coordinate system.   To avoid
confusion we refer to this first one as $SG_{zero}$.

\begin{figure*}
\centering                      \includegraphics[width=0.49\hsize]{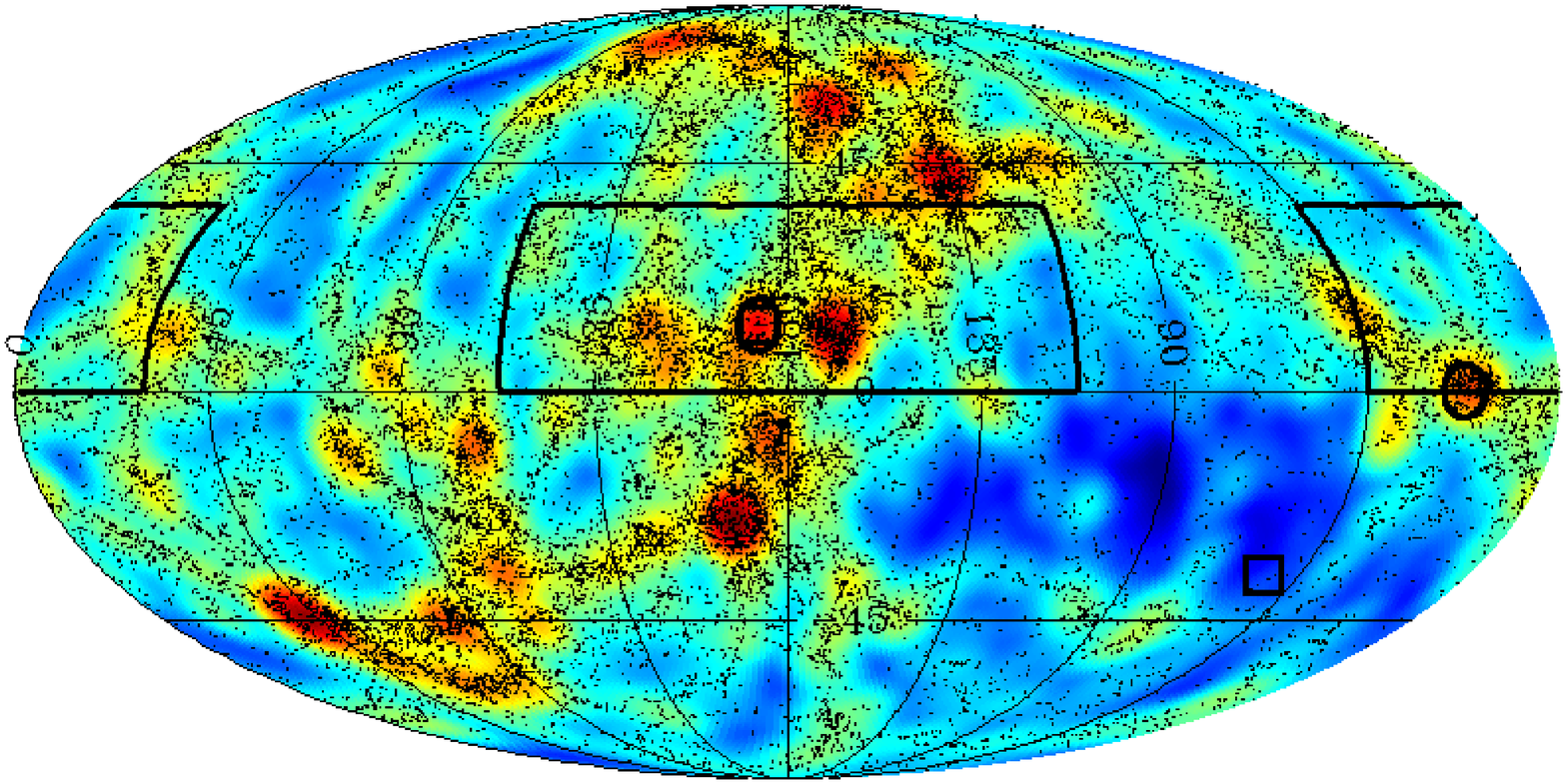}
\includegraphics[width=0.49\hsize]{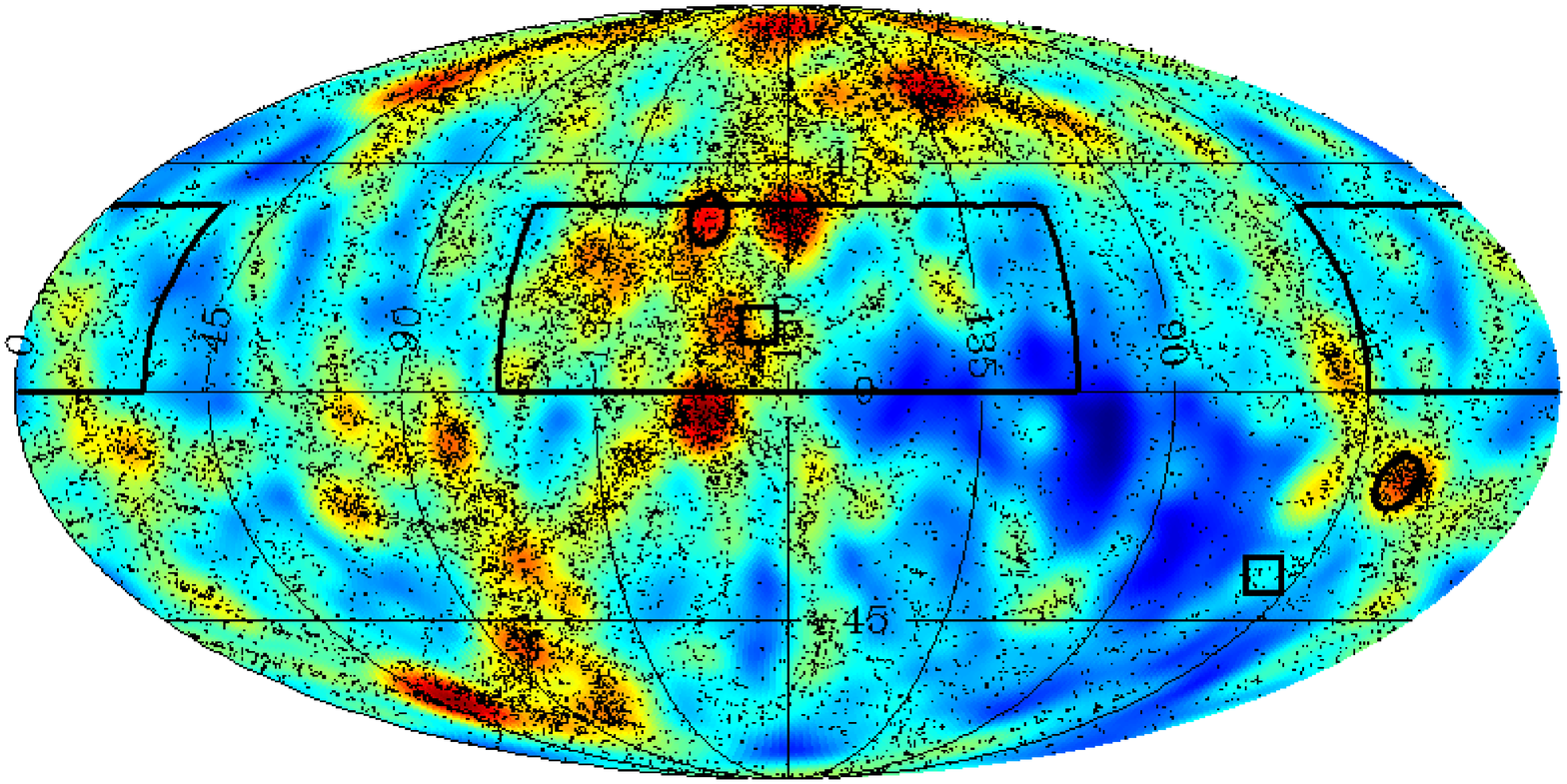}
\caption{\label{mollweide}{\it Left panel:} sky maps in equatorial coordinates
  (RA  and DEC) of  all halos  within a  sphere of  $20\hMpc$ centered  in the
  MW-like  halo   in  the   $\Lambda$CDM  simulation  for   coordinate  system
  $SG_{zero}$. The  color scale represents  a smoothed mass-weighted  count of
  the number of  halos per pixel, going  from blue to red for  lower to higher
  counts.  The  angular  positions  of  all  halos  with  masses  larger  than
  $5\times10^{9}h^{-1}M_{\odot}$  are displayed  with black  dots.   The boxes
  marked with solid black lines in the center and in the sides are the VdR and
  aVdR respectively.  The locations of the simulated Virgo and Fornax clusters
  appear as  black circles, the  correspondent locations of the  real clusters
  appear  as black  squares.   {\it Right  panel:}  same as  on  the left  for
  coordinate system $SG_{min}$.}
\end{figure*}

The result  of our coordinate  definition is visualized  in the left  panel of
Fig.~\ref{mollweide}.  It shows  a sky  map with  the angular  distribution of
halos in  Equatorial coordinates (RA, DEC)  in a Mollweide  projection for the
$\Lambda$CDM  simulation.   Only  halos   within  $20\hMpc$  from  the  origin
(simulated MW) are included. The size of  the sphere is limited by the size of
the simulation  box. Larger  radii are likely  to contain  spurious structures
caused by periodic boundary conditions.  The map was created using the HEALPIX
software   \footnote{http://healpix.jpl.nasa.gov/}   using  $N_{pix}=12(64)^2$
pixels.  The  value of  each pixel is  given by  a mass-weighted count  of all
halos located in that pixel. Afterwards, a smoothing of the map was done using
a  Gaussian beam  with a  FHWM of  $7\degr$; the  different color  scale  is a
measure of  the values  in the map,  from red-to-blue for  high-to-low values.
The   angular    positions   of   all   halos   with    masses   larger   than
$5\times10^{9}\hMsol$ also  appear in the map as  black points. The
voids  and high  density  regions within  the  local simulated  volume can  be
clearly appreciated in the sky  map.  The prominent high density region
crossing the  whole map almost vertically at  the center is  the simulated  LSC with
Virgo roughly in the middle  (black circle).  By construction, the location of
the real Virgo (black square) is  identical with the simulated one.  The boxes
in the center and  on the sides of the map give the  boundaries of the VdR and
aVdR, respectively, accessible to the ALFALFA survey.

Since the LSC is  roughly in place in our CSs and  because the simulated Virgo
and the real  one are at the same  angular position in the sky  (and almost at
the same distance from the LG), we believe that the procedure described in the
last paragraph led  to an appropriate coordinate system  allowing us to use
our CSs to simulate the VdR of the sky surveyed by ALFALFA.  However, a visual
impression of the aVdR  (enclosed by the boxes on the sides  of the sky map in
the left panel of Fig.~\ref{mollweide}) indicates tentative problems with this
simulated area  of the sky.  It contains a  significant part of  a filamentary
structure   with   a  high   density   of   halos,   stretching  from   around
($0\degr$,$0\degr$)  to ($45\degr$,$+20\degr$).  Such structure  is associated
with the simulated  Fornax cluster. This cluster was imposed as  a part of the
simulation constraints, see Table 1, it appears as a black circle in the right
corner of the sky map. Its angular  position is however out of place, the real
angular position  of the  Fornax cluster  is marked as  the lower  right black
square in Fig.~\ref{mollweide}.

Such  deviation is  within the  expected variations  of CSs  due  to intrinsic
uncertainties. For instance, the velocity  constraints used to produce the CSs
have  still  large  errors,  making  the  random component  of  the  CSs  more
significant.  Also, the present day positions of the massive clusters obtained
from  observational data are  imposed on  the initial  conditions of  the CSs.
Both of these conditions imply that the dynamical evolution of the simulations
shifts  the  positions of  the  clusters  and  modify the  structures  finally
obtained in the CSs.   Furthermore, since scales smaller than $\sim5\hMpc$
are  unconstrained, the  position of  the LG  is not  imposed directly  by the
constraints. Finally,  to be  able to explore  low mass  halos we have  used a
small box size for  the CSs; since the LSC stretches from  one side of the box
to  the  other,  periodic  boundary   conditions  distort  the  shape  of  the
LSC. Despite  all these difficulties,  we are confident  that the CSs  and the
choice we have made for the location of the LG are reliable enough to make the
comparison we intend in this work.

The result to keep  in mind is that Fornax appears inside  the aVdR whereas in
reality  it does not.  Then, by  using $SG_{zero}$  we would  over-predict the
abundance of halos in this region since by mistake our simulated survey probes
a region of higher density than the one expected from observations.

To ameliorate  this problem,  we carried out  an additional adjustment  of the
coordinate system.  We relax the  requirement that the angular position of the
simulated Virgo has to coincide exactly  with the angular position of the real
Virgo.  Instead,  we look for  a coordinate system  with the same  origin, and
that minimizes  the quadratic sum of  the distances between  the simulated and
real  clusters  Virgo  and Fornax.  We  refer  to  this coordinate  system  as
$SG_{min}$.  The distribution of halos after  this rotation can be seen in the
sky map at the right  panel of Fig.~\ref{mollweide}.  In the coordinate system
$SG_{min}$,  the distances  between the  real and  simulated Virgo  and Fornax
clusters are  $4.5\hMpc$ and $5\hMpc$  respectively.  The main effect  of this
rotation  is that  the (ALFALFA)  aVdR region  no longer  includes significant
parts of the filamentary structure associated to the Fornax cluster.
 
We are confident  that the adoption of the  new coordinate system, $SG_{min}$,
is   consistent  with   the  freedom   inherent  to   the  CSs.    In  section
~\ref{sec:fieldofview} we  investigate the  impact of the  rotation in  a more
general sense. There we can show  that minor rotations, as the one adopted for
the final  adjustment of the coordinate  system, change the  halo abundance in
the  VdR by  less than  30\%. Thus,  the adjustment  of the  coordinate system
introduces only a small uncertainty that  we will however keep in mind for the
interpretation of our results.  Here  we have only discussed the adjustment of
the $\Lambda$CDM  coordinate system,  we repeated the  same procedure  for the
$\Lambda$WDM simulation.

\section{Halo abundance}
\label{sec:CDMresults}
\subsection{Global mass function}
\label{sec:globalmass}
\begin{figure}
\centering              \includegraphics[height=14cm,width=14cm]{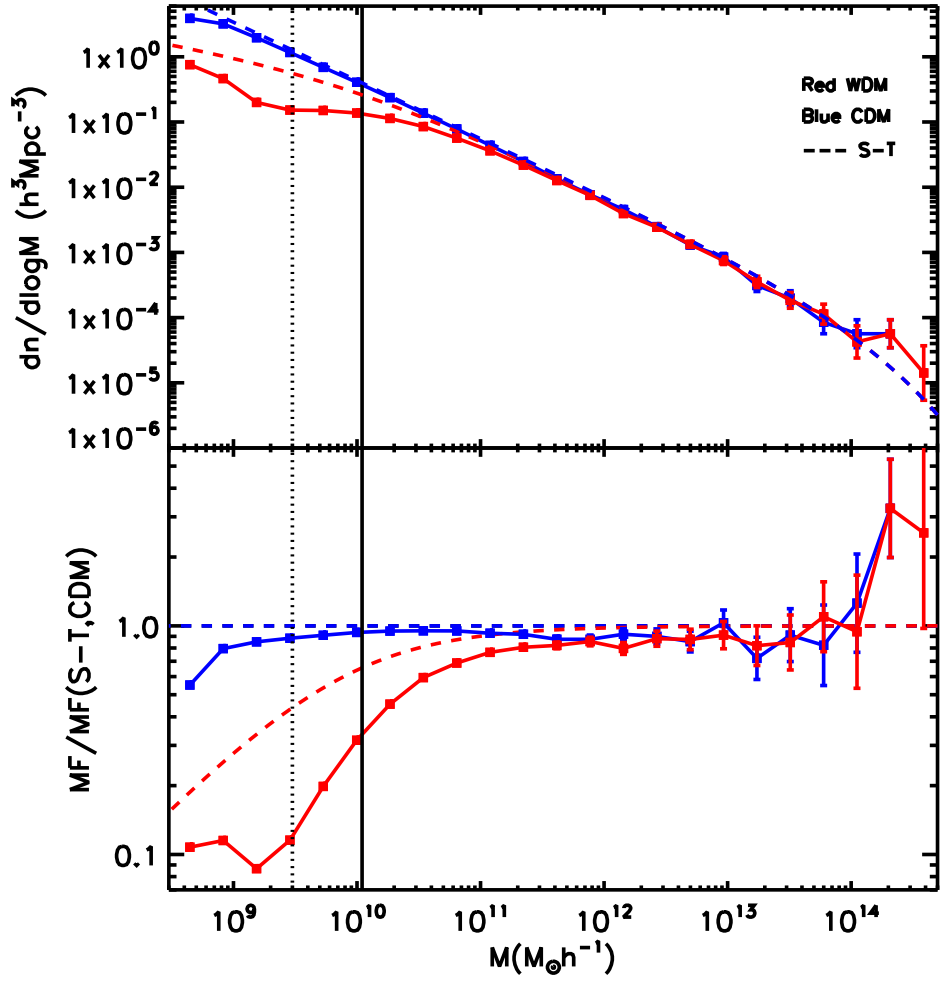}
\caption{{\it  Upper  panel:}  differential   MFs  for  the  $\Lambda$CDM  and
  $\Lambda$WDM   constrained   simulations   (blue   and   red   solid   lines
  respectively).   Analytical  predictions   following  the  Sheth  \&  Tormen
  formalism  for both  cosmologies  are  shown as  dashed  lines.  {\it  Lower
    panel:} ratio  of the MFs  to the value  of the MF  given by the  Sheth \&
  Tormen  formalism  for the  $\Lambda$CDM  cosmology.  The  solid and  dotted
  vertical lines  mark the values  of the filtering  and limiting mass  of the
  $\Lambda$WDM     simulation:     $1.1\times10^{10}h^{-1}$M$_{\odot}$     and
  $3\times10^{9}h^{-1}$M$_{\odot}$ respectively.}
\label{MF}
\end{figure}

%%
%%Fig. 4
%%
The blue and  red solid lines in the upper panel  of Fig.~\ref{MF} display the
differential  mass  function  for   the  $\Lambda$CDM  and  $\Lambda$WDM  CSs,
respectively.  Analytical predictions for both cosmologies are shown as dashed
lines  following   the  Sheth   \&  Tormen  formalism   \citep[S-T  formalism,
][]{Sheth-Tormen-99,   Sheth-Mo-Tormen-01,    Sheth-Tormen-02}.    For   their
computation  we used  the public  code described  in \cite{Reed-07}.   In this
prediction, the only difference between both cosmologies is the suppression of
the power spectrum at small  scales in the $\Lambda$WDM case.  The statistical
error  bars  presented  in  the  figure  are  Poisson  errors,  employing  the
definition given in \citet{Lukic-07}: $\sigma_{\pm}=\sqrt{N+1/4}\pm1/2$, where
$N$ is the number of halos per  bin.  The value of the filtering mass for the
WDM  simulation, $1.1\times10^{10}\hMsol$,  is marked  in  the figure
with a vertical solid line.
 
At the high mass  end there is an excess of massive  halos in our simulations.
This is caused  by the constraints which enforce the growth  of a very massive
structure, the LSC, within the  relatively small volume of the simulation box.
At  the   low  mass  end  we   see  discreteness  features   as  described  in
\cite{Wang-White-07}. According to their study,  the limiting mass that can be
trusted  is given  by the  formula: $M_{lim}=10.1\bar{\rho}  d k_{peak}^{-2}$,
where $d$  is the  interparticle separation,
$k_{peak}$  is the  wave number for  which $k^3P(k)$ reaches  its maximum and
$\bar{\rho}=\Omega_{DM}\rho_{crit}$.
For       the      case       of       our      $\Lambda$WDM       simulation:
$M_{lim}=3\times10^{9}\hMsol$.   As can  be seen  in Fig.~\ref{MF},
this value (indicated by the dotted  vertical line) marks the mass limit below
which there is an  artificial rise in the WDM MF, an  indication for the onset
of discreteness effects.

The difference between  the MF of the different  simulations becomes even more
clear in  the lower  panel of Fig.~\ref{MF},  where we  plot the ratio  of the
measured  MFs to  the value  of  the S-T  MF for  the $\Lambda$CDM  cosmology.
Clearly, the abundance  of low mass halos in  the $\Lambda$WDM case (for masses
larger than  $M_{lim}$) is considerably  lower than the predictions  using the
S-T     formalism.     This    discrepancy     has    been     found    before
\citep[e.g.,][]{Bode-Ostriker-Turok-01},  so   one  should  not   expect  good
agreement between the  S-T approach in the range of masses  close to and below
the $\Lambda$WDM filtering mass. In the following we will use the $\Lambda$CDM
S-T mass function as a reference to present some of our results.
\subsection{Global velocity function}
\label{sec:globalvelocity}
As  was  mentioned in  the  introduction, a  more  direct  comparison for  the
abundance of structure between our  CSs and observations of the local Universe
can be achieved  by constructing the velocity function  of dark matter halos.
The maximum rotational velocities are  used instead of the masses to calculate
the  abundance  of halos  per  logarithmic  $V_{max}$  bin.  Provided  that  a
physical  connection between  $V_{max}$  and the  measured maximum  rotational
velocity of spiral  galaxies can be established, our  velocity function can be
directly  compared with  observations of  galactic discs.   Further  below, in
\ref{sec:galaxies}, we describe a simplified model which is designed to
accomplish this goal.

%%
%%Fig. 5
%%

\begin{figure}
\centering \includegraphics[height=14cm,width=14cm]{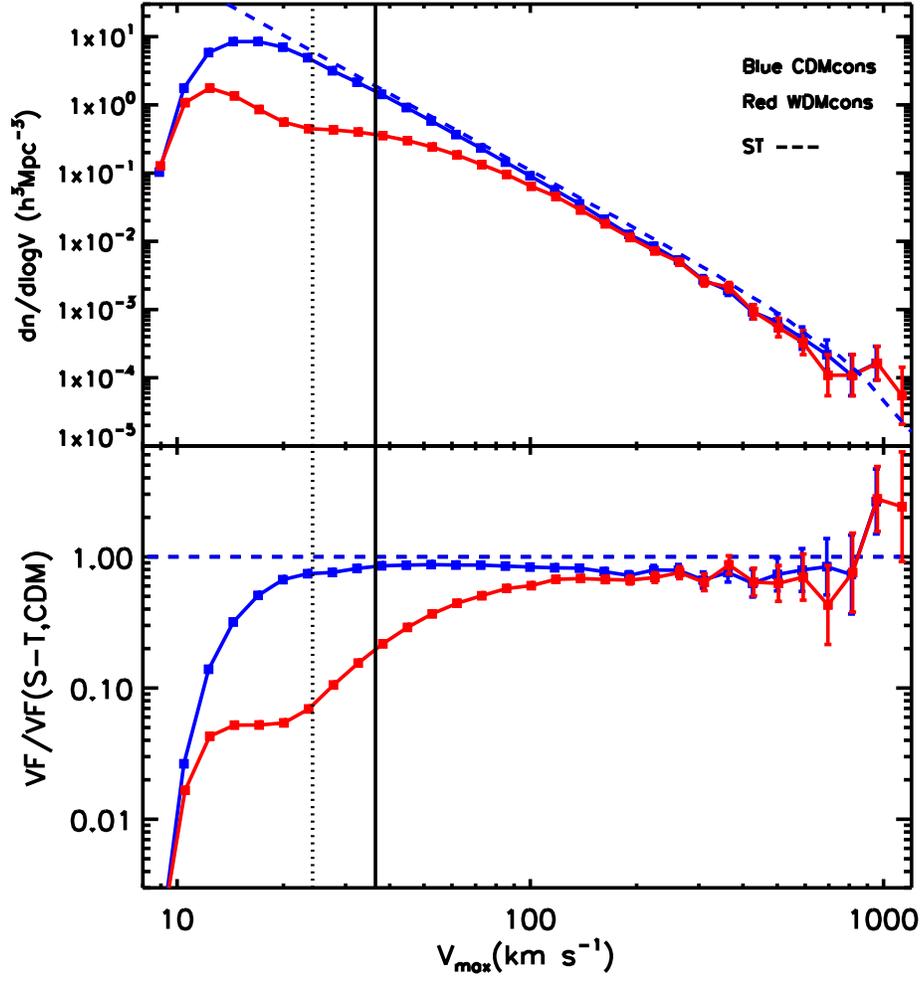}
\caption{The same  as Fig.~\ref{MF} but  for the VFs  instead of the  MFs.  In
  this case no analytical prediction  is shown for the $\Lambda$WDM cosmology.
  The vertical lines indicate the maximum velocities corresponding to the
  filtering  and  limiting  masses,  with  values  of  $36\kms$  and  $24\kms$
  respectively.}
\label{VF}
\end{figure}

The upper panel  of Fig.~\ref{VF} shows the differential VF  for halos in both
simulations   (the   line   styles   and   colors   are   the   same   as   in
Fig.~\ref{MF}). Also shown is a prediction for the VF in the $\Lambda$CDM case
(dashed line)  obtained using the procedure outlined  in \citet{Sigad-00}.  In
brief, the  procedure is the  following: i) a  ``virtual'' sample of  halos is
generated in such a way that its MF mimics the one given by the S-T formalism;
ii) a concentration value taken from a log-normal distribution \citep{Jing-00}
is  assigned  to  each  halo;  mean  and standard  deviation  values  for  the
log-normal    distribution     were    taken    from     the    analysis    by
\citet{Maccio-Dutton-vandenBosch-08}  for the  set of  simulations  with WMAP3
cosmology:   $\langle$log$c\rangle=1.775-0.088$log$M_{200}$  and  $\sigma_{log
  c}=0.132$; iii) assuming  a NFW profile, we compute  $V_{max}$ for each halo
using the concentration and virial  velocity corresponding to that halo (e.g.,
see eq. 7  of \citet{Sigad-00}). A prediction in the  case of the $\Lambda$WDM
simulation is not given since the previously described steps are all uncertain
in this case.

The vertical solid and dotted  lines in Fig.~\ref{VF} are the estimated values
for the maximum velocities corresponding  to the filtering and limiting masses
for  the $\Lambda$WDM simulation.   For the  computation of  the corresponding
virial velocities we used the relation between the virial mass of the halo and
its  virial radius:  $M_{vir}=\frac{4}{3}\pi\Delta\rho_{crit}r_{vir}^3$.  Then, the virial velocity is simply given by,
$V_{vir}^2=GM_{vir}/r_{vir}$.   Using  the  same  mass-concentration  relation
described in the  paragraph above, we compute the  maximum circular velocities
for  the filtering  and limiting  mass  and find  the values  of $36\kms$  and
$24\kms$.

The lower  panel of Fig.~\ref{VF} shows the  ratio of the measured  VFs to the
analytical  one displayed as  dashed line  in the  upper panel\footnote{Recall
  that the  analytical result is given  by the S-T  formalism for $\Lambda$CDM
  only}. This  figure is analogous to  the ratios of  the MFs as shown  in the
lower  panel of  Fig.\ref{MF}.  The  analytical  estimate for  the VF  follows
closely the result of the $\Lambda$CDM CS for most of the velocity range.  The
difference  at the high  mass end  is due  to the  constraints imposed  in the
simulation.

The downward bend  for the $\Lambda$CDM velocity function  at the low-velocity
end     is      caused     by      the     mass     cut-off      for     halos
($\sim4\times10^{8}\hMsol$). The  reason why we do not  see an abrupt
cut off  is due to  the spread in  halo concentrations which causes  a similar
spread in $V_{max}$.  The behavior of  the $\Lambda$WDM VF is analogous to the
behavior  of  the  $\Lambda$WDM  MF.  Discreteness effects  can  be  seen  for
velocities lower  than the limiting  $V_{max}$ (dotted line).   For velocities
just above  this limiting  value, the difference  between both  simulations is
approximately an order of magnitude.
\subsection{Abundance of halos in the local environment}
\label{sec:abundnace}
%%
%%Fig.6 
%%
We  now turn our  attention from  the global  to the  local abundance  of dark
matter  halos.   In  Fig.~\ref{MF_ratio_radius}  we  show  the  ratio  of  the
differential MFs to  the $\Lambda$CDM S-T prediction for  spheres of different
size centered on the LG.  The upper  and lower panels show the results for the
$\Lambda$CDM and $\Lambda$WDM simulations.  In both panels the ratios decrease
with increasing  radius of the spheres.   The smallest sphere has  a radius of
$15\hMpc$ resulting in the highest abundance (uppermost curve in each panel).
The following sequence  of curves corresponds to radii  increased by intervals
of $5\hMpc$ up to a maximum radius of $40\hMpc$.  The curve corresponding to a
sphere with radius  $20\hMpc$ is highlighted with a  thicker dash-dotted line.
For reference, the result for the MF  in the whole cubic box is presented as a
thick solid line (see Fig.~\ref{MF}). For spheres growing beyond the border of
the simulation box  we use periodic boundary conditions to  fill in the volume
of the sphere.

According  to  Fig.~\ref{MF_ratio_radius},  the  local  environment  shows  an
overabundance  of halos  compared  to the  mean  abundance in  the whole  box.
Within    $20\hMpc$   (which    was   the    radius   used    to   produce
Fig.~\ref{mollweide}) the halo abundance is  about 2 times larger than that in
the entire box.   In what follows all results are restricted  to a sphere with
radius  of $20\hMpc$,  which is  close  to the  maximum radius  that a  sphere
centered at  the LG can  have and still  lie completely inside  the simulation
box.
\section{Predictions for the ALFALFA survey}
\subsection{Simulated field of view of ALFALFA}
\label{sec:fieldofview}
%%
%%Fig. 7
%%
Fig.~\ref{SG_20_CDM_WDM}  displays  the  distributions  of  halos  within  the
ALFALFA field of view projected onto the plane of the supergalactic coordinate
system,    $SG_{min}$,    which    was    introduced    at    the    end    of
section~\ref{sec:Coordinates}. The panels on the left are for the $\Lambda$CDM
simulation and on the right for the $\Lambda$WDM simulation.  The red and blue
dots give the position of halos within the VdR and the aVdR respectively. Only
halos with  a distance to  the LG less  than $20\hMpc$ and with  masses larger
than  $5\times10^{9}\hMsol$  are  shown.   The difference  in  halo
abundance between both regions is  clearly visible. 

Fig. \ref{Halo_count}  shows the  radial dependence of  the number  density of
halos, $n_h$, normalized  to the total number density  of halos, $n_{sph}$, in
the $20\hMpc$ sphere.  The red and blue  lines show the result for the VdR and
aVdR, and the black line for the  whole sphere. The upper and lower panels are
for the $\Lambda$CDM and $\Lambda$WDM simulations, respectively. The peak at a
distance of $\sim11\hMpc$ in the VdR  (red line) is caused by halos associated
with the  LSC in  the vicinity of  the Virgo  cluster.  For radii  larger than
$10\hMpc$ the VdR  is significantly overdense, compared to  the density in the
sphere, whereas the aVdR is underdense at all radii.

%%
%%Fig.8
%%

The    difference    between    VdR    and    aVdR   is    also    clear    in
Fig.~\ref{VF_ratio_ALFALFA}  where  we  show  the  differential  VF  for  both
simulations (solid  lines for $\Lambda$CDM and dotted  lines for $\Lambda$WDM)
in different  regions around the LG:  the whole $20\hMpc$  sphere (black solid
and dotted lines), the VdR (red  lines) and the aVdR (blue lines).  The values
for the  filtering (vertical solid  line) and limiting (vertical  dotted line)
velocities for the $\Lambda$WDM simulation are also shown for reference.

%%
%%Fig.9
%%

In both  simulations, the  difference between  the VdR and  aVdR is  clear and
qualitatively as expected, the former being an overdense region and the latter
an underdense region, relative to the whole region contained within the sphere
of $20\hMpc$  radius. We note however,  that the aVdR has  actually a similar,
although smaller,  density than the mean cosmological  density, especially for
the $\Lambda$CDM simulation. This becomes apparent from the comparison between
the  blue solid  and black  dashed  lines representing  the aVdR  and the  S-T
prediction, respectively, in the  $\Lambda$CDM case (see also Fig.  \ref{VF}).
This  fact  is  partially related  to  the  limitations  of our  CSs  (already
mentioned in detail  in section 3).  For halo  velocities lower than $100\kms$
and larger than  the WDM limiting velocity ($24\kms$),  the difference between
the VdR and  aVdR is approximately constant in both  cosmologies, the ratio of
their differential VFs in this range is roughly 3.

%%
%%Fig.10
%%

To  conclude this  section we  investigate the  robustness of  our  results in
relation  to  the  choice  of   the  coordinate  system.   For  that  purpose,
Fig.~\ref{impact_coo} displays a comparison  of the VFs for moderate rotations
of    the    supergalactic    coordinate    system    defined    in    section
\ref{sec:Coordinates}. The comparison is done  using the ratio of the velocity
functions  to the  velocity function  in the  whole sphere  with $20\hMpc$
radius ($VF/VF_{20}$).  The black and blue dashed lines show the ratios of the
VFs for the VdR and aVdR  based on the initial coordinate system, $SG_{zero}$.
This coordinate system  was set up to optimize the  agreement between the real
and simulated Virgo  location, but failed to get the  Fornax cluster in place.
In  this  system,  the simulated  aVdR  contains  a  significant part  of  the
filamentary structure  associated with the  Fornax cluster (see left  panel of
Fig.~\ref{mollweide}), which  is not being  surveyed in the observed  field of
view.  Therefore, the abundance of halos in the simulated aVdR is unexpectedly
high.   The  solid  black  and  blue  lines show  the  same  quantities  using
$SG_{min}$. With it, the simulated aVdR does not contain Fornax any longer and
the halo abundance decreases substantially.   The results based on the systems
$SG_{zero}$  and  $SG_{min}$  differ  by  $\sim15-20\%$ for  the  VdR  and  by
$\sim40-50\%$  for the aVdR.   The red  region shown  in Fig.~\ref{impact_coo}
displays the  ratio of the  VFs functions in  the VdR for rotations  of system
$SG_{zero}$  up to  $20\degr$ in  both supergalactic  latitude  and longitude.
This figure indicates that the volume density of halos in the VdR stays within
$20-30\%$  of its  value in  the  coordinate system  $SG_{zero}$ for  moderate
rotations around it  and halo velocities below $100\kms$ (which is the
range we are  ultimately interested in).  The velocity function  in the VdR is
therefore robust  against moderate rotations  of the coordinate system  due to
the appearance  of adjacent high  and low density  regions. For the  aVdR, the
results  are   more  sensitive  to   rotations  of  the   original  coordinate
system. These general results for the choice of coordinate system apply in the
$\Lambda$WDM case.
\subsection{Velocity function of disk galaxies}
\label{sec:galaxies}
In  order to directly  compare our  simulations with  observations we  need to
populate the dark  matter halos with galaxies.  Essentially, we  need to use a
method to  connect the  maximum circular velocity  measured for  disk galaxies
with  the  properties of  the  hosting  halo.   Since a  full  semi-analytical
treatment goes  beyond the  scope of  the current study,  we use  a simplified
scheme  that, nevertheless,  allows us  to  make predictions  on the  velocity
function of galaxies in the  local environment.  First, appropriate halos need
to  be selected.   The high-velocity  (large-mass) halos  are  associated with
groups and  clusters of galaxies. In  our scheme we exclude  halos with masses
larger  than $10^{13}\hMsol$  since we  are  interested in  low mass  isolated
galaxies. Next, we assume each of the remaining halos to contain only one disk
galaxy (for  isolated galaxies in the  local Universe, the  fraction of spiral
galaxies                 lies                 between                 80-90\%,
e.g. \citealt{Sulentic-06,Hernandez-Toledo-08}).  Although many of these halos
contain significant  substructures, which in  principle can be  populated with
satellite  galaxies, we  restrict  the present  study  to main  halos and  the
central galaxies  within them. This can  be justified because  the fraction of
satellite  galaxies lies  between  $10-40\%$ \citep{Zheng-Coil-Zehavi-07}  and
approximately  $50\%$ of  them are  unlikely to  be detected  in  surveys like
ALFALFA.   The  latter estimate  is  roughly  correct  because, for  optically
selected  samples, a  large fraction  of  satellite galaxies  have red  colors
\citep{Weinmann-06,Wang-07,Font-08}  and have  probably already  lost  most of
their gas.   HI studies of nearby  groups \citep[e.g.][]{Kilborn-05} similarly
show  that  about half  of  the satellites  would  lay  below the  sensitivity
threshold  for HI  detection.  Therefore,  our results  may  underestimate the
abundance of HI  sources by $5-20\%$, and thus, the lack  of satellites in our
modelling is not a major source of uncertainty.

We compute  the circular velocities of the  disks lying at the  center of each
halo  using the  analytical model  of \citet{Mo-Mao-White-98}.   The principal
hypotheses of this model are  the conservation of specific angular momentum of
both, the dark matter and the gaseous components, and the equality of disk and
halo specific angular momenta ($J_{disk}/M_{disk}=J_h/M_h$) during the process
of disk formation.  These assumptions  are key conditions for the formation of
realistic disks in hydrodynamical simulations \citep{Zavala-Okamoto-Frenk-08}.
The rotation curve of the galactic system (disk+halo) is given by the combined
gravitational effects of  the disk (which ends up  with an exponential surface
density  profile)   and  the  adiabatically  contracted   halo.   The  maximum
rotational  velocity  ($V_{max}$:  disk+halo)  of  the disk  and  the  maximum
rotational velocity of the pure  dark matter halo ($V_{max,h}$) are related by
$V_{max}=G(\lambda,f_{disk})V_{max,h}$,   the  function  $G(\lambda,f_{disk})$
depends on  the spin  parameter of  the halo, $\lambda$,  and the  fraction of
baryonic mass that  is used to assemble the  disk, $f_{disk}=M_d/M_{vir}$. The
$V_{max}/V_{max,h}$  ratio  increases   with  $f_{disk}$  and  decreases  with
$\lambda$.  We  note  that  this  ratio  is nearly  independent  of  the  halo
concentration; for fixed values of $f_{disk}$ and $\lambda$, variations of $c$
within the typical scatter in the mass-concentration relation produce a change
of less than $2\%$.   The function $G(\lambda,f_{disk})$ has been approximated
in \citet{Zavala-Th} with a fitting  function that has an accuracy larger than
$96\%$ for $\lambda\in[0.02,0.1]$:
\begin{equation}\label{G}
G(\lambda,f_{disk})=1.04\left(1-\frac{0.11f_{disk}+5\times10^{-4}}{\lambda}\right)^{-1}
\end{equation}
The parameter  $f_{disk}$ actually depends  on the processes  occurring during
galaxy formation, such  as gas cooling and feedback. For  simplicity we take a
constant value $f_{disk}=0.03$  for all galaxies.  However, at  the end of the
following section  we address  the effects of  SN feedback.  For  the constant
value of $f_{disk}$,  we take $\lambda_{lim}=0.02$ as a  limiting value, below
which  the disk would  be unstable  \citep{Mo-Mao-White-98}.  We  note however
that the large majority of  the halos in our CSs have $\lambda>\lambda_{lim}$,
halos with lower values are not considered in the analysis.

%%
%%Fig.11
%%

Applying the  scheme described  above we obtain  the velocity function  of the
modeled  disk galaxies.  The result  is shown  in Fig.~\ref{VF_ALFALFA_disks},
colors    and   line    styles   are    the    same   as    those   used    in
Fig.~\ref{VF_ratio_ALFALFA}. We have included in  the figure the values of the
filtering  and  limiting velocities  for  halos  related  to the  $\Lambda$WDM
(vertical  solid and dotted  lines). These  values mark  lower limits  for the
corresponding  quantities in  the case  of modeled  disks in  the $\Lambda$WDM
scenario.
\subsection{Comparison with the early ALFALFA catalog release}
\label{sec:Comparison}

\subsubsection{Sample selection and corrections to the line-width W$_{50}$}

By the  end of  the writing of  this paper,  only three catalogs  have been
publicly   released   by  the   ALFALFA   collaboration,   two   in  the   VdR
\citep{Giovanelli-07,Kent-08}    and    the   other    one    in   the    aVdR
\citep{Saintonge-08}.  These catalogs comprise only
$\sim6\%$ of the final volume. The
first  two cover an area from  $11^h30^m\lesssim  R.A.\lesssim14^h$   and  $+08\degr<  DEC
<+16\degr$  and  the  last  one  from $22^h<  R.A.<03^h$  and  $+26\degr<  DEC
<+28\degr$.  Despite  the limited  volume, we make  an attempt to  compare our
results to the observational data released so far.

For such comparison, we take a  sample of the ALFALFA sources according to the
following criteria: i)  distances lower than $20\hMpc$; ii)  exclusion of High
Velocity  Clouds (HVCs)  and  sources with  no  measurement of  HI mass;  iii)
removal   of   sources  with   no   inclination   measurement   or  with $i\leq30\degr$
\footnote{Velocity measurements are not reliable for $i\leq30\degr$ because of
  the  large deprojection  correction and  the uncertainty  associated  to the
  estimate of $i$.  We have tested the influence of  different values for this
  lower   limit,  0$^{\circ}$,  45$^{\circ}$,   60$^{\circ}$,  and   found  no
  significant impact  in our  results.}; and finally,  iv) removal  of sources
showing clear signs of interaction within a projected angular distance smaller
than the beam size of the Arecibo antenna ($\sim3.5'$).

Criterion i) is the  most stringent of all, only 14$\%$ of  the sources in the
three original catalogs fulfill it.  Of the remaining galaxies, 81$\%$ satisfy
criteria ii)-iv).  The final sample consists of 186 galaxies in the VdR and 15
in the aVdR.

It  is necessary to  obtain measurements  for the  inclination of  the sources
since the 21cm line-width $W_{50}$ (measured  at the $50\%$ peak level) can be
associated  with $V_{max}$ only  after appropriate  corrections, one  of them,
deprojection to  an edge-on view.  We  discuss further below  how we corrected
the values of $W_{50}$.

To get the inclinations, we  extracted the minor-to-major axis ratios (b/a) by
cross      checking      the     sources      with      the     GOLD      Mine
Database\footnote{http://goldmine.mib.infn.it/}     \citep{Gavazzi-03},    the
Cornell             HI              Archive             of             pointed
sources\footnote{http://arecibo.tc.cornell.edu/hiarchive/}
\citep{Springob-05},     and    the    NASA/IPAC     Extragalactic    Database
(NED)\footnote{http://nedwww.ipac.caltech.edu/}.   Using the  axis  ratios, we
compute the inclinations with the formula:
\begin{equation}\label{incl}
\cos^2(i)=\frac{(b/a)^2-q_0^2}{1-q_0^2}
\end{equation}
where $q_0$  is the intrinsic axial ratio  of a galaxy seen  edge-on, we adopt
$q_0=0.2$ as  a fiducial value for  all galaxies in our  sample, independent
of morphological type \citep[e.g.][]{Tully-09}. For criterion
iv)  we checked,  whenever  it was  possible,  the optical  counterpart of  the
sources using  the Sloan Digital  Sky Survey (SDSS) website\footnote{http://www.sdss.org}.

The value of $W_{50}$ given in the raw catalogs has already been corrected for
instrumental  broadening.  We further  corrected  these  values for  turbulent
motions  and  for  inclination   effects  following  the  procedure  given  in
\citet{Verheijen-Sancisi-01}: 
\begin{eqnarray}
W_{50}^2=\frac{1}{\rm{sin}(i)}\left[W_{50,R}^2+W_{t,50}^2\left(1-2e^{-\left(\frac{W_{50,R}}{W_{c,50}}\right)^2}\right)-\right.\nonumber\\
\left.2W_{50,R}W_{t,50}\left(1-e^{-\left(\frac{W_{50,R}}{W_{c,50}}\right)^2}\right)\right]
\end{eqnarray}
where  $W_{50,R}$ is  the  raw value  corrected  for instrumental  broadening,
$W_{t,50}=5\kms$ and $W_{c,50}=100\kms$. After these corrections, the maximum
rotational velocity of the galaxies  can be estimated with reasonable accuracy
as $W_{50}/2$.   We make  however the following  comment on the  estimation of
$V_{max}$ using  $W_{50}$: although the HI line-width  provides no information
on the radial rotation profile of the galaxy, most of the HI gas is located in
the   outer   part   of   disk   galaxies  (typically   beyond   three   scale
lengths)\footnote{We note  that strongly HI deficient low  mass galaxies, like
  those inside  the Virgo cluster, are  typically not detected  by the ALFALFA
  survey.}.   For  large disks,  $W_{50}$  provides  a  measure of  rotational
velocity in  the regions where  the rotation curve  is already flat  or rising
slowly
\citep[e.g.][]{Catinella-Giovanelli-Haynes-06,Catinella-Haynes-Giovanelli-07}.
In  the case  of galaxies  with lower  velocities ($<75\kms$),  their rotation
curves  are  typically  thought  to  be  still rising  to  the  last  measured
point. However,  \citet{Swaters-09} have recently analyzed in  detail a sample
of  dwarf galaxies  and found  that the  shapes of  their rotation  curves are
similar to those of more  massive galaxies; in particular, the rotation curves
typically start  to flatten at two  disk scale lengths. If  these results hold
for the galaxies  in our sample, then  we can be confident to  use $W_{50}$ to
get $V_{max}$ without an important systematic underestimation.
%%
%%Fig. 12
%%

\subsubsection{The HI velocity function}

The  angular   position  of  the  final   sample  of  galaxies   is  shown  in
Fig.~\ref{ALFALFA_sources_limited}  using  equatorial  coordinates (upper  and
middle panels for the VdR and aVdR respectively). As a reference, the position
of M87  is marked in the  figure with a red  star.  The lower  panel shows the
number density  of sources  as a function  of their  distance to the  MW.  The
prominent  overdensity  around 11$\hMpc$  is  caused  by  galaxies in  the
vicinity of the  Virgo cluster.  A similar feature is present  in our CSs (see
Fig.~\ref{Halo_count}).  We  note that distances in the  ALFALFA catalogs take
into  account  the   peculiar  velocity  field  according  to   the  model  of
\citet{Tonry-00}.

%%
%%Fig.13
%%

The HI  velocity function of  the sample  of galaxies in  the VdR is  shown in
Fig.~\ref{VF_limited_virgo} with red square  symbols. The values were computed
using   the   $\Sigma(1/\mathbb{V}_{max})$   weighting  method   proposed   by
\citet{Schmidt-68}.  $\mathbb{V}_{max}$ is  the  volume given  by the  maximum
distance,  $D_{max}$,  where a  given  source could  be  placed  and still  be
detected by the  survey. This distance is determined  by the sensitivity limit
of  the survey,  given in  the ALFALFA  survey design  for a  signal  to noise
threshold of 6 (see eq.  5 of \citealt{Giovanelli-05b}):
\begin{equation}\label{limit}
\left(\frac{D_{max}}{\rm Mpc}\right)^2=\frac{1}{0.49}f_{\beta}t_s^{1/2}\left(\frac{\rm{M_{HI}}}{10^6\Msol}\right)
\left(\frac{W_{50}}{200\kms}\right)^{\gamma}
\end{equation}
where $M_{HI}$  is the  HI mass associated  to the source,  $\gamma=-1/2$ for
$W_{50}<200\kms$  and  $\gamma=-1$   for  $W_{50}\ge200\kms$.   The  parameter
$f_{\beta}$  quantifies  the fraction  of  the  source  flux detected  by  the
telescope's beam,  for simplicity, we treat  all sources as  point sources and
take $f_{\beta}=1$. Since the beam size  of the Arecibo antenna is $3.5'$, the
majority of the sources can be treated  in this way.  We take a fiducial value
of 48~s for  the integration time $t_s$.  For  comparison with our simulations
we are restricting  the analysis to distances lower  than $20\hMpc$, thus,
effectively:
$\mathbb{V}_{max}=\textrm{min}(\mathbb{V}(D_{max}),\mathbb{V}(20h^{-1}\textrm{Mpc}))$.
For  the  majority  of the  sources,  $D_{max}>20\textrm{h}^{-1}\textrm{Mpc}$,
therefore,  the  volume-weights  have  only  a  minor  impact  in  the  number
count. However, three sources in the lower velocity bins have a very low value
of  $V_{max}$ and their  weights deviate  strongly from  the average  in their
respective   bins;  we  removed   them  since   they  are   not  statistically
representative  and  can lead  to  a  strong  overestimation of  the  velocity
function.

The results from our CSs for the corresponding field of view are also shown in
Fig.~\ref{VF_limited_virgo}.  The  red dashed  and dotted areas  encompass the
1$\sigma$  regions,  using Poisson  statistics,  for  the  predictions in  the
$\Lambda$CDM and  $\Lambda$WDM cases, respectively.  The  VFs were constructed
taking  into account  the  sensitivity limit  of  the survey,  using the  same
$\Sigma(1/\mathbb{V}_{max})$  weighting  method as  for  the observations  and
renormalizing  the result  according to  the  fraction of  galaxies that  were
excluded from the final observational  sample due to the $30\degr$ inclination
cutoff.  To  properly apply  the sensitivity  limit we would  need to  give an
estimate of the gas fraction, $f_{gas}=M_{HI}/M_{disk}$, in the modeled disks,
that depends  on the efficiency of  gas transformation into stars  and the gas
infall history. For simplicity we take $f_{gas}=1$, but we note that the value
of $f_{gas}$ is irrelevant for  our particular analysis, see discussion by the
end of the section.  The sensitivity  limit has a relevant effect only for the
low velocity disks.

For velocities larger  than $\sim80\kms$, the VF of  both CSs match reasonably
well the  observational data .   This result is  not trivial, for  example, at
$V_{max}=100\kms$, the value of the VF for halos in the whole simulated box is
approximately an  order of  magnitude lower than  the value associated  to the
observed sample  in the VdR.  Therefore, we  confirm that the CSs  are able to
simulate properly the overdense VdR.  For velocities in the range 35-80$\kms$,
the $\Lambda$CDM  simulation overpredicts the value of  the velocity function,
increasingly  for  lower velocities.   On  the  other  hand, the  $\Lambda$WDM
simulation is in  good agreement with the observed  data, with values slightly
higher.   The  low  velocity   end,  $V_{max}<35\kms$,  is  not  suitable  for
comparison since both,  simulations and observations are not  complete for the
lower velocities. The  minimum halo mass in our  simulations that is reliable,
according to a  comparison of the MF with theoretical  expectations in the low
mass  end  (see  Fig.~\ref{MF}),  sets  a  lower  limit  of  completeness  for
$V_{max}$.  For the  $\Lambda$CDM simulation,  this mass  is $\sim10^9\hMsol$,
corresponding to $V_{max}\sim24\kms$; in the $\Lambda$WDM case is given by the
limiting mass $3\times10^9\hMsol$  related to $V_{max}\sim29\kms$. The typical
sensitivity limit for the ALFALFA  survey goes down to $M_{HI}=10^7\hMsol$ for
distances up to $20\hMpc$, see eq.  (\ref{limit}), a source with this mass can
have a $V_{max}$ value within a broad range, due to the natural scatter on the
$V_{max}-M_{HI}$  relation.  An examination  of our  galaxy sample  shows that
this range  is $\sim13-35\kms$, which puts  a lower limit  of $\sim35\kms$ for
the  completeness of  the  sample. Taking  the  theoretical and  observational
limits into  account, a consistent comparison  can only be done  for values of
$V_{max}$ above $\sim35\kms$.

So  far  we have  used  a constant  disk  baryon  fraction $f_{disk}=0.03$  to
populate the dark  matter halos with galaxy disks.   A more realistic approach
would be to  include the effects of SN feedback  in $f_{disk}$. This inclusion
would only have an important effect  for low velocity halos if the gas outflow
produced by SN is sufficient to deplete  the forming galaxy of gas and push it
below the sensitivity  limit of ALFALFA. We now  incorporate this effect using
the disk  galaxy evolutionary  model of \citet{Dutton-vandenBosch-08}  for the
case where the outflow of gas is energy driven: the kinetic energy of the wind
is a fraction  $\epsilon_{EFB}=0.25$ of the kinetic energy  produced by SN. In
this case, $f_{disk}$ becomes a strong  function of halo mass (see upper panel
of Fig.  8 in \citealt{Dutton-vandenBosch-08}).  We adopt the fitting function
obtained by  the authors for the  energy driven feedback model  (see their eq.
43  and  table  3)  and  apply  it  to  our  model.   The  results  appear  in
Fig.~\ref{VF_limited_virgo} as  dashed and  dotted lines for  the $\Lambda$CDM
and $\Lambda$WDM CSs.  The net impact is a marginal reduction of the VF at the
low velocity  end.  A more  drastic reduction would  be possible with  a model
that significantly lowers the value of $f_{gas}$, which we have set to one for
simplicity.  For  instance, the  lowest  value  we  have set  for  comparison,
$V_{max}=35\kms$, is  related on average  to a halo  of $6\times10^{9}\hMsol$,
which in  the case  of the SN  feedback model corresponds  to $f_{disk}=0.02$,
thus the modeled disk would lie below the sensitivity limit of the survey only
if  $f_{gas}$  is lower  than  0.08.   In our  scheme  the  gas fraction  only
determines if a given halo should be counted or not, thus, the previous result
indicates  that the  specific  value of  $f_{gas}$  has no  impact unless  the
galaxies  we are  aiming  to compare  with  have very  low  gas fractions.  HI
deficient low mass galaxies are  typically present inside galaxy clusters like
Virgo  and  are  unlikely  to  be  detected by  ALFALFA;  they  also  have  no
counterpart in  our simulations  since we have  dealt exclusively  with halos,
excluding the subhalos within them.  Galaxies  in the field are less likely to
have such low gas fractions.  Nevertheless, we explicitly checked that this is
indeed the case  for the sample of  galaxies in the VdR. For  that purpose, we
identified the SDSS optical counterparts of the HI sources and used the method
described in \citet{Blanton-Roweis-07} to  estimate the stellar masses of 73\%
of the galaxies in the final VdR  sample.  We found that the majority of these
galaxies  have  large gas  fractions,  approximately  90\%  of the  ones  with
$V_{max}<100\kms$ have $f_{gas}>0.1$.

%%
%%Fig.14
%%

In  Fig.~\ref{VF_limited_anti}  we  show   analogous  results  for  the  aVdR.
Although in this  case the statistical significance of  the observed sample is
much lower (only 15 sources), the main results observed in the case of the VdR
are reproduced,  indicating that simulations  are able to probe  properly this
region as well.
\section{Discussion and Conclusions}
\label{sec:Discussion}

N-body simulations with constrained initial conditions, set-up to resemble the
spatial distribution of dark matter in the local Universe, are a powerful tool
to make detailed  comparisons between predictions of the  dark matter paradigm
and observational evidence in our local environment.

Using    the   algorithm    of   constrained    realizations    developed   by
\citet{Hoffman-Ribak-91}     and     the      approach     laid     out     in
\citet{Kravtsov-Klypin-Hoffman-02} and  \citet{Klypin-03}, we have  run a pair
of  CSs that incorporate  nearby observational  data sets  as input  for their
initial conditions. One simulation follows the evolution of structure in a CDM
model and the other one is based on a thermal WDM particle with a mass of 1keV
resulting in an effective filtering mass of $\sim10^{10}\hMsol$.

After an appropriate choice of the coordinate system, the simulations are able
to  reproduce  the  overall  spatial  distribution  of  the  most  significant
structures within 20$\hMpc$ of the Local Group, namely, the LSC, including
the  Virgo cluster,  as well  as  the Fornax  cluster laying  in the  opposite
direction (see section 3).

The mass  and velocity  functions of  halos in the  whole simulated  boxes are
consistent  with  theoretical  expectations.   The $\Lambda$CDM  case  follows
closely  the estimates  from  the Sheth  \&  Tormen formalism,  except in  the
highest  mass end  due to  the  influence of  the LSC.   For the  $\Lambda$WDM
cosmogony,  the results  lie close  to the  $\Lambda$CDM case  for  halos with
masses higher than the filtering mass; for lower masses, the mass and velocity
functions flatten  and then rise  due to spurious numerical  fragmentation for
masses  $\lesssim3\times10^9\hMsol$ confirming the  limiting mass
formula  given by  \citet{Wang-White-07}.   The mass  resolution  of the  CSs,
$m_{DM}=1.63\times10^7\hMsol$,   allows  us  to derive robust
results for halos with masses larger than this limiting mass, corresponding to
maximum rotation velocities of 24$\kms$.

\begin{figure}
\centering                  \includegraphics[height=14cm,width=14cm]{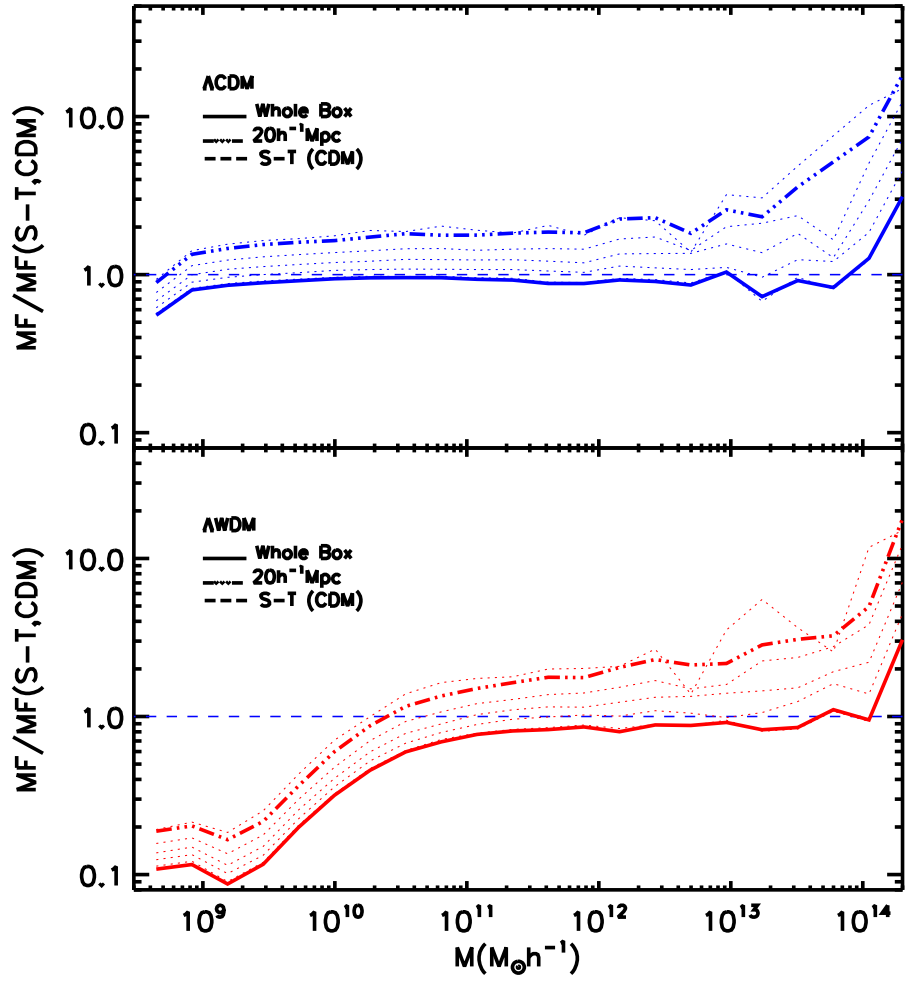}
\caption{Ratio of the differential MFs  for the $\Lambda$CDM (upper panel) and
  $\Lambda$WDM (lower  panel) CSs to the  prediction of the  S-T formalism for
  the $\Lambda$CDM  cosmology.  The solid  lines represent the result  for the
  whole simulation boxes  (see Fig.~\ref{MF}).  The blue dashed  lines are for
  the  S-T prediction.  The  other lines  are for  spheres of  different radii
  centered  at  the  LG.   The  curve  for  the  $20\hMpc$  radius  sphere  is
  highlighted with a thick dash-dotted line.}
\label{MF_ratio_radius}
\end{figure}

\begin{figure}
\centering                  
\scalebox{0.9}{\includegraphics*[2.0in,1.0in][7.0in,8.5in]{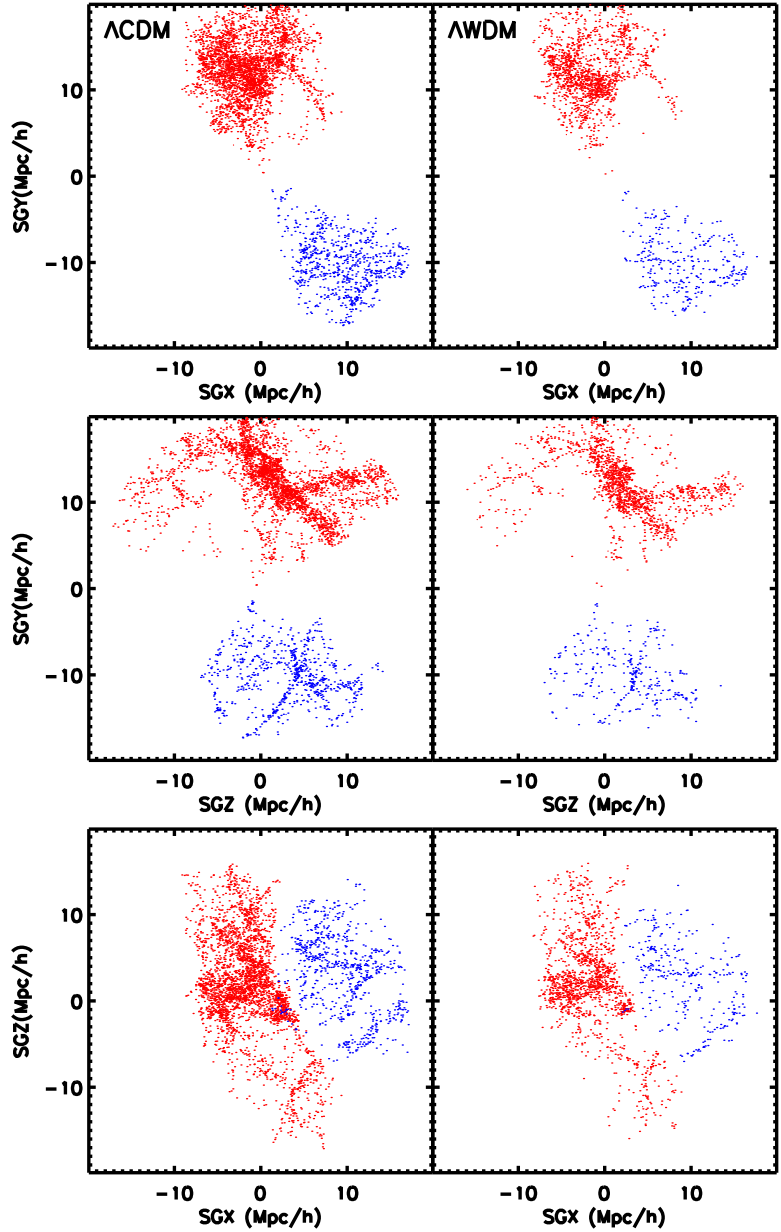}}
\caption{
  \label{SG_20_CDM_WDM} 
  Projections of the halo distributions in  the ALFALFA field of view onto the
  planes of the supergalactic coordinate system $SG_{min}$.  Only halos within
  $20\hMpc$    from   the   MW    halo   and    with   masses    larger   than
  $5\times10^{9}h^{-1}M_{\odot}$ are  shown.  The red and  blue dots represent
  halos inside the  VdR and aVdR respectively.  The left  and right panels are
  for the $\Lambda$CDM and $\Lambda$WDM simulations, respectively.}
\end{figure}

\begin{figure}
\centering                  \includegraphics[height=14cm,width=14cm]{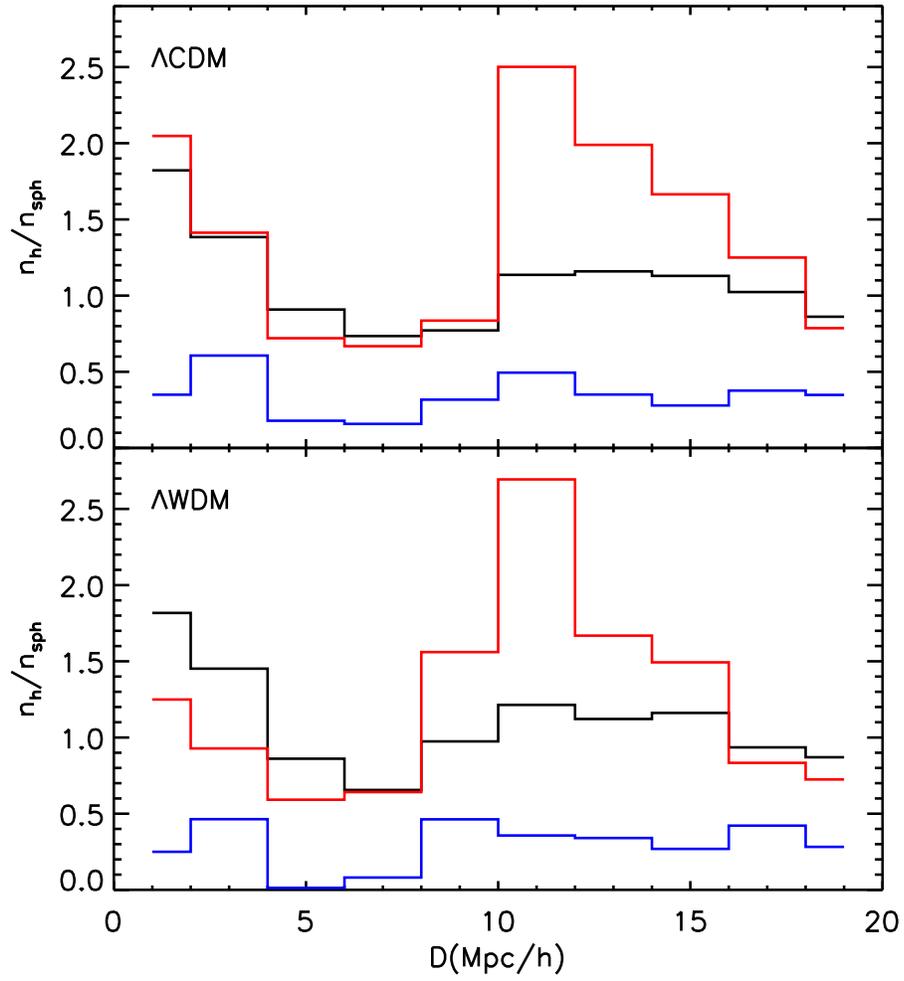}
\caption{{\it Upper panel:} number density  of halos, in the $\Lambda$CDM
  simulation, as a function of distance  to the LG for
  the halos displayed in  Fig.~\ref{SG_20_CDM_WDM}.  The values are normalized
  to the number  density within the $20\hMpc$ sphere.  Red  and blue lines are
  for the VdR  and aVdR respectively.  The black line  gives the radial number
  density profile for the whole sphere.{\it Lower panel:} same as on the left
  for the $\Lambda$WDM simulation.}
\label{Halo_count}
\end{figure}

For  the local  environment, the  general prediction  of the  CSs is  that the
region  within 20$\hMpc$ of  the LG  is overdense  by a  factor of
$\sim2$   compared   to  the   MF   of   the   whole  simulated   boxes   (see
Fig.~\ref{MF_ratio_radius}).  Such  result goes in  the same direction  as a
recent  result  reported   by  \citet{Tikhonov-Klypin-08},  showing  that  the
luminosity function in a sample of galaxies within 8$\hMpc$ is larger than
the universal luminosity function by a factor of 1.4.

Since we have  shown that our CSs are capable of  reproducing the abundance of
halos in  the local environment,  we have obtained  predictions for the  VF of
halos in  the field of view,  within 20$\hMpc$, that is  being surveyed by
ALFALFA.   The VF  has the  important advantage  over the  MF that  it  can be
compared more directly with observational  data since it avoids the problem of
relating halo masses to galaxy luminosities.

\begin{figure}
\centering                  \includegraphics[height=14cm,width=14cm]{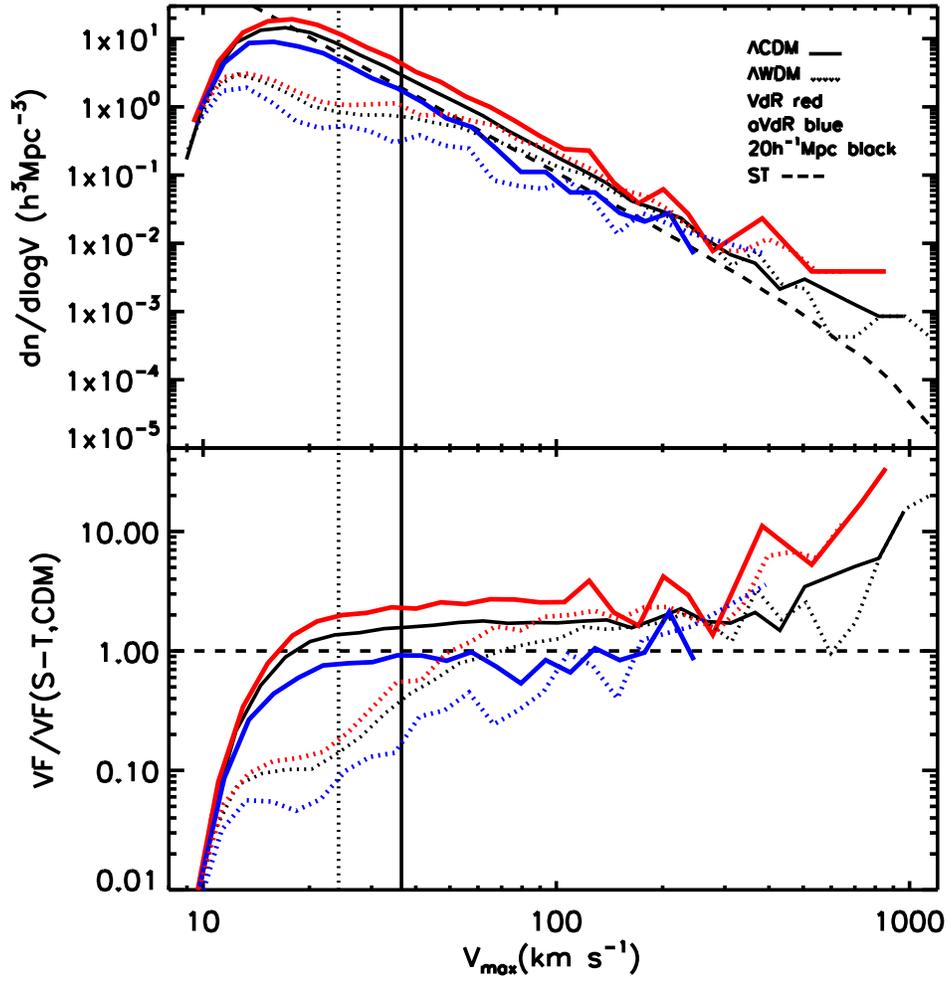}
\caption{{\it  Upper  panel:} differential  VFs  for  the $\Lambda$CDM  (solid
  lines) and  $\Lambda$WDM (dotted lines)  constrained simulations in  the VdR
  (red)  and aVdR  (blue). The  black (solid  and dotted)  curves are  for the
  sphere of $20\hMpc$ radius centered at the LG.  The S-T prediction, only for
  $\Lambda$CDM  cosmology, is  shown  as  a black  dashed  curve.  {\it  Lower
    panel:} ratio of  the VFs for the  simulations to the VF given  by the S-T
  formalism for the  $\Lambda$CDM cosmology (same line styles  as in the upper
  panel).   The  solid  and dotted  vertical  lines  mark  the values  of  the
  filtering  and limiting  velocities  of the  $\Lambda$WDM  simulation as  in
  previous figures.}
\label{VF_ratio_ALFALFA}
\end{figure}

\begin{figure}
\centering                  \includegraphics[height=14cm,width=14cm]{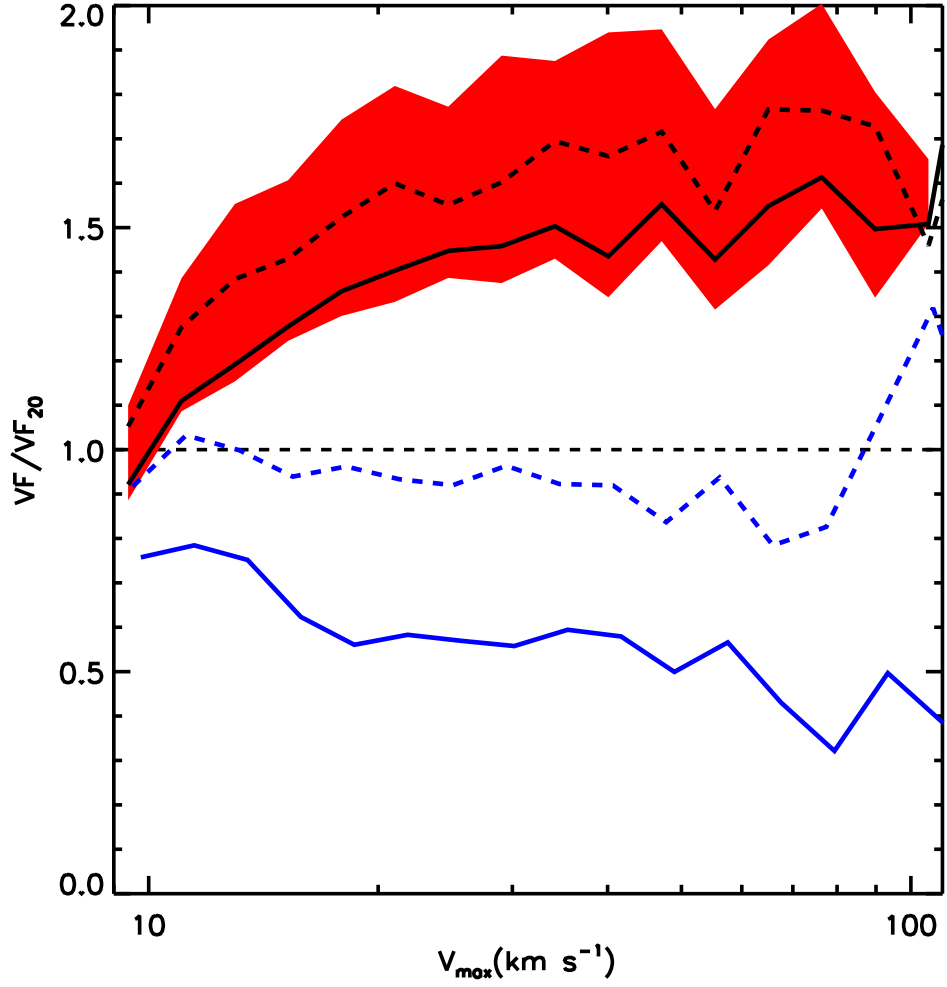}
\caption{Ratio of the  differential VF of the $\Lambda$CDM  simulation for the
  VdR to  the VF in  a sphere  of $20\hMpc$ radius. The red zone encompasses
  the value of these ratios for different
  rotations of  the supergalactic coordinate  system.  The results for  two of
  these  rotations,  corresponding  to the coordinates  systems  $SG_{zero}$  and
  $SG_{min}$  defined  in section  \ref{sec:Coordinates},  are highlighted  as
  black dashed and solid lines respectively.  Also shown are the ratios in the
  aVdR for these two rotations (blue lines).}
\label{impact_coo}
\end{figure}

\begin{figure}
\centering                  \includegraphics[height=14cm,width=14cm]{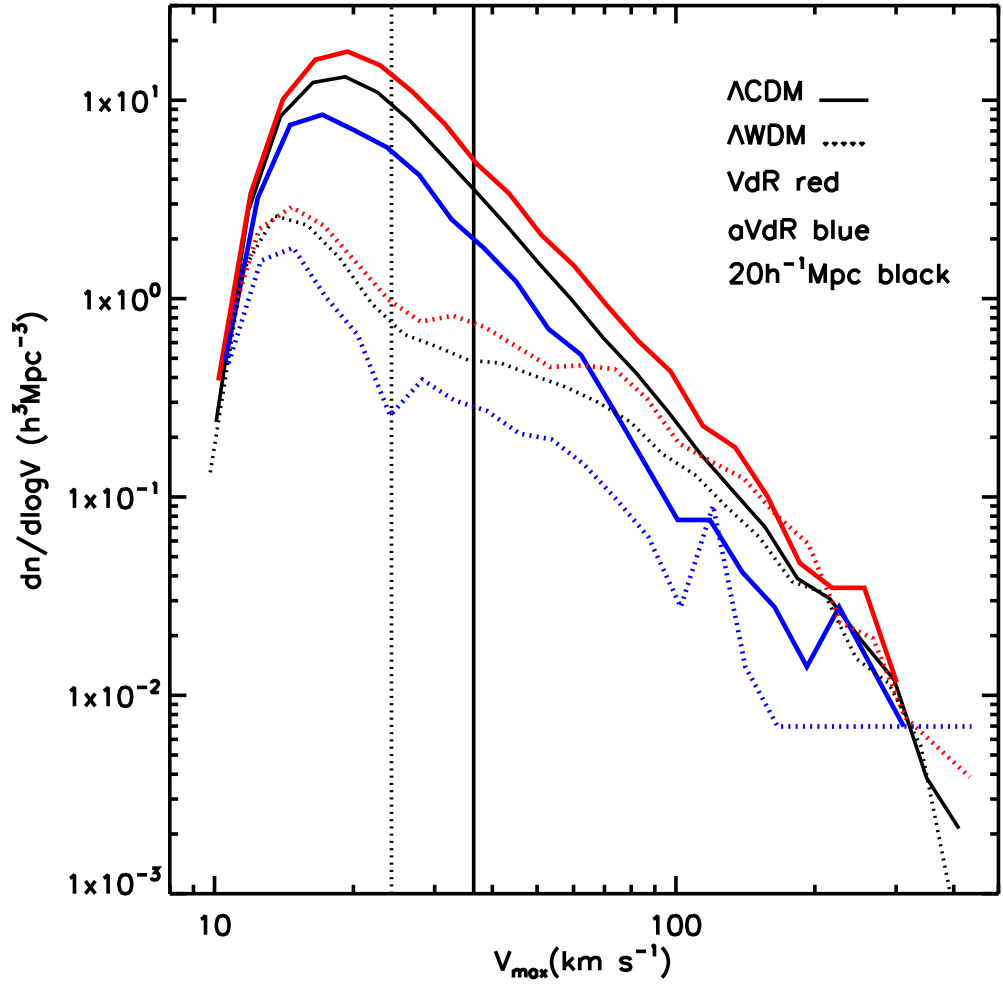}
\caption{Differential VFs of modeled galaxies (disk+halo).   Line-styles and correspondences are as
  in Fig.~\ref{VF_ratio_ALFALFA}.}
\label{VF_ALFALFA_disks}
\end{figure}

\begin{figure}
\centering                  \includegraphics[height=14cm,width=14cm]{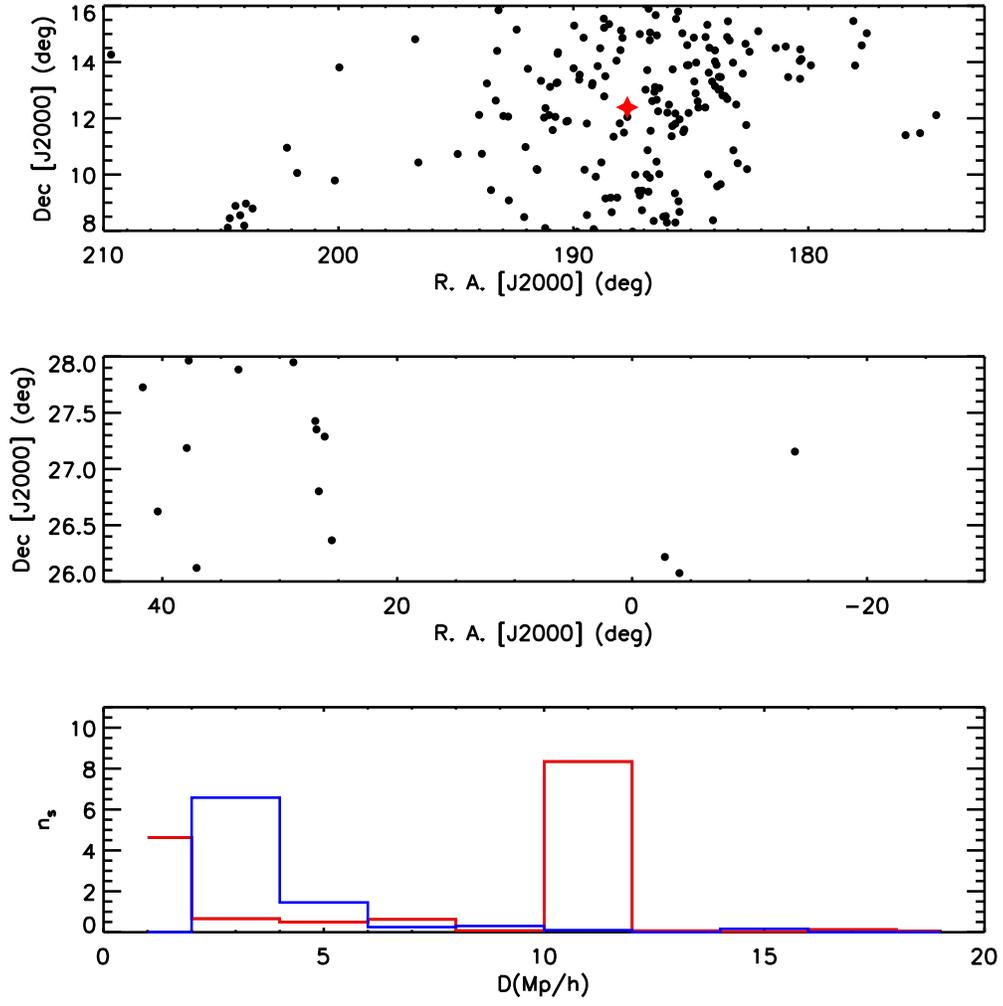}
\caption{{\it  Upper panel:  }distribution  in equatorial  coordinates of  the
  sample of HI sources taken from  the ALFALFA public catalogs released so far
  in the VdR.  The position of $M87$ in  the Virgo cluster is marked  as a red
  star. {\it Middle panel:  }sample of sources in the  aVdR.  {\it Lower panel:
  }the number  density of sources  as a function  of their distance to  the MW
  (red and blue for the VdR and aVdR respectively).}
\label{ALFALFA_sources_limited}
\end{figure}

\begin{figure}
\centering                  \includegraphics[height=14cm,width=14cm]{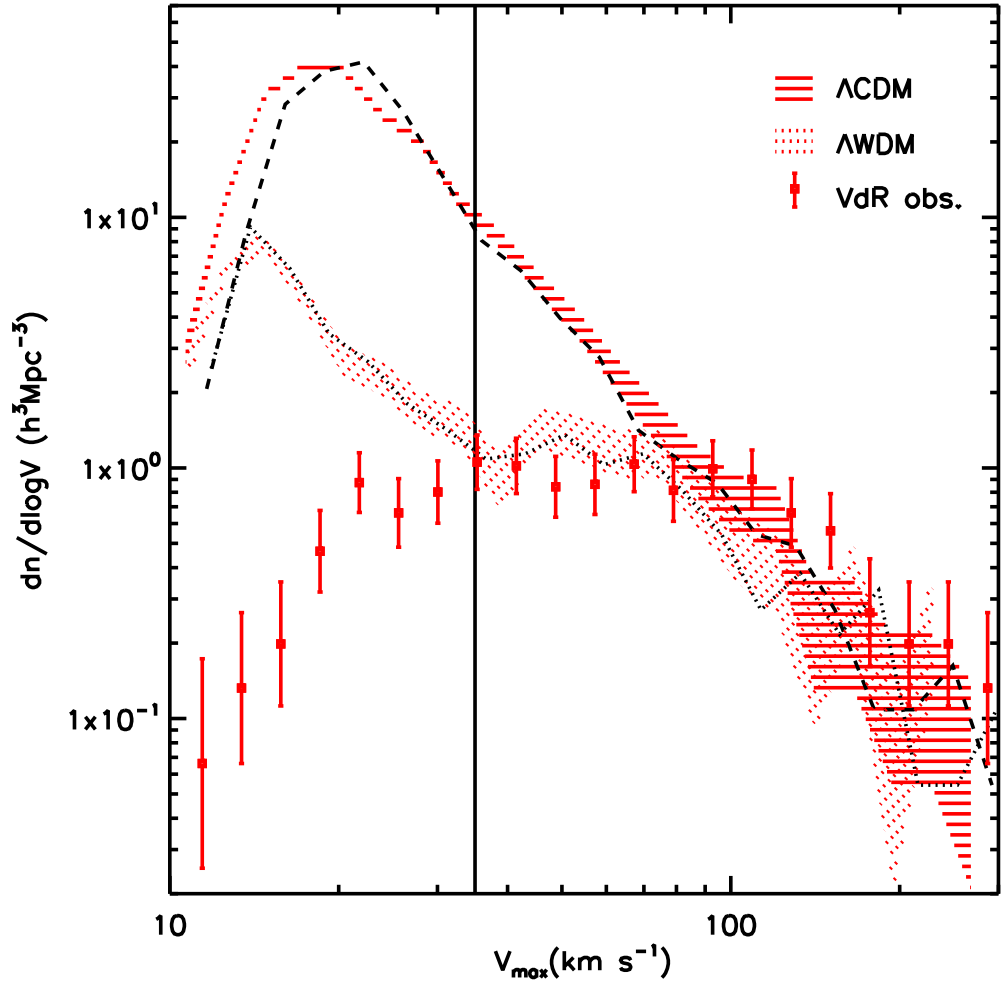}
\caption{VF for the sample of galaxies in the VdR taken taken from the ALFALFA
  catalogs (square symbols with error  bars).  Predictions from our CSs, using
  $f_{disk}=0.03$,  for  the observed  field  of  view  appear as  the  dashed
  ($\Lambda$CDM)  and dotted  ($\Lambda$WDM) red  areas, delimited  by Poisson
  error bars. The dashed and dotted  lines are predictions using a model where
  $f_{disk}$  is a  function  of halo  mass  incorporating the  effects of  SN
  feedback.   The sensitivity limit  of the  survey has  been included  in the
  results. The vertical solid line marks the value of $V_{max}$ down to which
  the simulations and observations are both complete.}
\label{VF_limited_virgo}
\end{figure}

\begin{figure}
\centering                  \includegraphics[height=14cm,width=14cm]{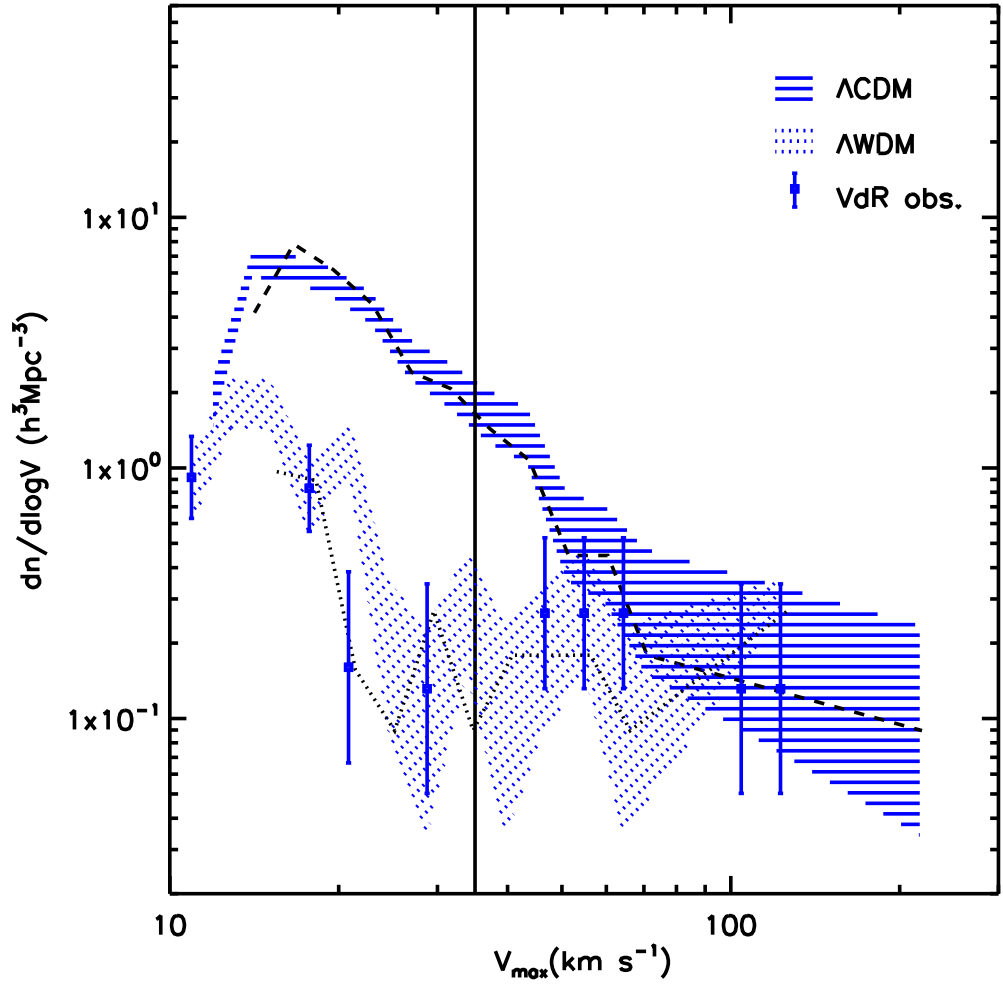}
\caption{The  same  as Fig.  \ref{VF_limited_virgo}  but  for the  limited
  survey coverage of the aVdR.}
\label{VF_limited_anti}
\end{figure}

After  completion,  the  ALFALFA   survey  will  detect  HI  sources  covering
7000deg$^2$  of the  sky  in two  different  regions. In  this  work, we  have
referred  to these  regions,  additionally constrained  to  $20\hMpc$, as  the
Virgo-direction region,  VdR, and  the anti-Virgo-direction region,  aVdR. Our
CSs predict  that the VF  of halos in  the VdR exceeds the  universal velocity
function  by  a factor  of  $\sim3$.  The VF  in  the  aVdR  is only  slightly
underdense     ($\sim10\%$)     than      the     universal     value     (see
Fig.~\ref{VF_ratio_ALFALFA}).

We have used  a simplified model to populate our halos  with disk galaxies. It
only incorporates the dynamical effects of the disks.  With this model, we are
able  to predict  the  VF of  disk galaxies  in  the two  regions explored  by
ALFALFA.  Although  the  survey is  not  complete  yet  we have  compared  our
predictions and the results from a  sample of galaxies taken from the catalogs
released so far, which cover only $6\%$ of the total planned volume.

For velocities in the range  between $80\kms$ and $300\kms$, the VFs predicted
for the $\Lambda$CDM and $\Lambda$WDM simulations agree quite well with the VF
of  the  sample  of  galaxies.   For  the VdR,  this  result  is  particularly
encouraging and reassures  the confidence in our CSs  to properly simulate the
local  environment:  despite  the   small  volume  used  for  comparison,  the
simulations are able  to predict the shape and normalization of  the VF in the
high velocity regime. In the VdR, the normalization is an order of
magnitude larger than for the universal VF.

For  velocities larger  than the  minimum mass  we can  trust  for comparison,
$\sim35\kms$ ,  and lower  than $80\kms$, the  predictions agree well  for the
$\Lambda$WDM  cosmogony   (contrary  to  recent  claims,  see   section  9  of
\citealt{Blanton-Geha-West-08}), with the VF  being approximately flat in this
regime.   The  $\Lambda$CDM model,  however,  predicts  a  steep rise  in  the
velocity function towards low velocities; for $V_{max}\sim35\kms$, it predicts
$\sim10$ times  more sources than  the ones observed.   Using the same  set of
simulations Tikhonov  et al. (2009),  in preparation, found that  the observed
spectrum  of mini-voids  in the  local volume  is in  good agreement  with the
$\Lambda$WDM  model  but  can  hardly  be explained  within  the  $\Lambda$CDM
scenario.

Although we have  only explored a simplified model to  populate our halos with
disk galaxies,  our results indicate  a potential problem of  the $\Lambda$CDM
paradigm in the low-velocity regime of dwarf galaxies. Nevertheless, there are
several issues that need to be addressed before reaching a strong conclusion.

On the observational  side, the sample of galaxies  we have analyzed comprises
only 186 galaxies in the VdR. It is of key importance to obtain the VF for the
complete volume  of ALFALFA and check  if the flattening at  low velocities is
reproduced. Another important issue is  to determine to which extent the value
of $W_{50}$ is representative of the maximum rotational velocity for the least
massive galaxies.  A recent  analysis by \citet{Swaters-09} indicates that the
rotation  curves for late-type  dwarf galaxies  are similar  to those  of more
massive  galaxies,  starting to  flatten  at  two  disk scale  lengths,  thus,
$W_{50}$ would  not underestimate  strongly the value  of $V_{max}$  for these
galaxies since the HI gas typically extends beyond three disk scale lengths.

From a  theoretical perspective, astrophysical  phenomena such as  SN feedback
and UV  photoionization play an  important role to  deplete gas from  low mass
halos.   The relevant  influence of  these effects  on the  VF comes  from the
typical sensitivity limit  of the survey, allowing to  detect HI masses larger
than $10^7\hMsol$ at distances $D\le20\hMpc$.   We have explored the former of
these effects  by using  the model presented  in \citet{Dutton-vandenBosch-08}
and  found  little  impact  on  the  VF.   Regarding  UV  background  heating,
\citet{Hoeft-06}  found   that  this  mechanism  is  not   very  efficient  in
evaporating all  baryons from  dwarf sized halos  below a  characteristic mass
scale  of $6\times10^9\hMsol$ ($V_{max}\sim35\kms$).   Since according  to our
findings, a strong suppression of  gas is needed for halos with characteristic
velocities  $V_{max}<60\kms$   to  flatten   the  velocity  function   in  the
$\Lambda$CDM case,  it is unlikely  that UV photoionization could  account for
it.

\acknowledgments

The  simulations   used  in  this   work  were  performed  at   the  Barcelona
Supercomputing Centre  (BSC), the Leibniz  Rechenzentrum Munich (LRZ)  and the
Shanghai Supercomputer  Center. The cpu  time used at  BSC and LRZ  was partly
granted by the DEISA Extreme  Computing Project (DECI) SIMU-LU.  JZ would like
to thank Volker Springel for  helpful comments in running the simulations, and
Alexander  Knebe  and Steffen  Knollman  for  help with  AHF.  JZ  and AF  are
supported by the Joint Postdoctoral  Program in Astrophysical Cosmology of the
Max  Planck   Institute  for   Astrophysics  and  the   Shanghai  Astronomical
Observatory. JZ  was partially  supported by the  CAS Research  Fellowship for
International Young Researchers. YPJ is supported by NSFC (10533030, 10821302,
10878001), by the Knowledge Innovation  Program of CAS (No. KJCX2-YW-T05), and
by  973 Program  (No. 2007CB815402).  Our collaboration  was supported  by the
ASTROSIM network  of the ESF. This  research has been supported  by the Israel
Science Foundation (13/08  at the HU). GY acknowledges  support of the Spanish
Ministry   of   Education    through   research   grants   FPA2006-01105   and
AYA2006-15492-C03.  Some of the results  in this paper have been derived using
the HEALPiX \citep{Gorski-05} package.  This research has made use of the GOLD
Mine Database and the NASA/IPAC Extragalactic Database (NED) which is operated
by the  Jet Propulsion Laboratory,  California Institute of  Technology, under
contract with the National Aeronautics and Space Administration. This research
has made  use of the  SDSS archive. Its  full acknowledgment can be  found at
http://www.sdss.org.

\bibliography{./lit}

%%%%%%Figures

%%Fig.1

%%Fig.2

%%Fig.3

%%Fig.4

%%Fig.5

%%Fig.6

%%Fig.7

%%Fig.8

%%Fig.9

%%Fig.10

%%Fig.11

%%Fig.12

%%Fig.13

%%Fig.14

%%%Table 1

\begin{table*}
 \centering
 \begin{minipage}{90mm}
  \caption{Main properties of the LG, Virgo and Fornax. The first line is the
    mass of the object and the second the relative distance to the
    LG. Distances are given in $h^{-1}$Mpc and masses in $10^{12}h^{-1}$M$_{\odot}$.}
  \begin{tabular}{@{}lrrrr@{}}
  \hline
   Object/Case & LG(MW+M31)\footnote{The distance is in this case the relative
   distance between MW and M31} & Virgo & Fornax \\
   \hline
   $\Lambda$CDM  & $1$  & $100$ & $27$\\
   & $0.9$  & $10.8$  & $10.5$\\
   $\Lambda$WDM  & $1$ & $94$ & $31$\\
   & $0.5$  & $10.2$  & $10.6$\\
   Obs.\footnote{Cluster values according to \citet{Girardi-98}}  &
   1.6\footnote{Central 
     values of the masses estimates according to \citet{Xue-08} for the MW and from
     \citet{Evans-Wilkinson-00} for M31} & 204 & 31\\
   & 0.6 & 11.4 & 15.0\\
   \hline
  \end{tabular}
 \end{minipage}
\end{table*}

\end{document}